\documentclass[review]{elsarticle}

\usepackage{lineno, hyperref}
\usepackage{listings}
\modulolinenumbers[5]
\usepackage{amsmath,amssymb,amsfonts}
\usepackage{graphicx}
\usepackage{textcomp}
\usepackage{xcolor}
\usepackage{afterpage}
\usepackage{longtable}

\usepackage{verbatim}
\usepackage{booktabs} 
\usepackage{fancyhdr,graphicx,amsmath,amssymb}
\usepackage[ruled, vlined, linesnumbered]{algorithm2e}
\usepackage{enumerate}
\usepackage[utf8]{inputenc}
\usepackage{etoolbox}
\usepackage{ltablex,booktabs}
\usepackage{pdflscape}
\usepackage{tabulary}
\usepackage{float}
\restylefloat{table}
\usepackage{tabularx}
\usepackage{parskip}
\usepackage{adjustbox}
\usepackage{booktabs}

\lstdefinelanguage{Gherkin}{
	morekeywords = {
		Given,
		When,
		Then,
		And,
		Scenario,
		Feature,
		But,
		Background,
		Scenario Outline,
		Examples
	},
	sensitive=true,
	morecomment=[l]{\#},
	morestring=[b]",
	morestring=[b]'
}

\journal{Journal of Systems and Software}

\begin{document}
\begin{frontmatter}

\title{Behaviour Driven Development: A Systematic Mapping Study}


\author[mymainaddress]{Leonard Peter Binamungu \corref{mycorrespondingauthor}}
\ead{lepebina@udsm.ac.tz}

\author[mymainaddress]{Salome Maro \corref{mycorrespondingauthor}}
\cortext[mycorrespondingauthor]{Corresponding author}
\ead{salomehonest@gmail.com}

\address[mymainaddress]{Department of Computer Science and Engineering, College of Information and Communication Technologies, University of Dar es Salaam, Dar es Salaam, Tanzania}

\begin{abstract}
 \textit{Context:} Behaviour Driven Development (BDD) uses scenarios written in semi-structured natural language to express software requirements in a way that can be understood by all stakeholders. The resulting natural language specifications can also be executed to reveal correct and problematic parts of a software. Although BDD was introduced about two decades ago, there is a lack of secondary studies in peer-reviewed scientific literature, making it difficult to understand the state of BDD research and existing gaps.  

 \textit{Objective:} To understand the current state of BDD research by conducting a systematic mapping study that covers studies published from 2006 (when BDD was introduced) to 2021. 

 \textit{Method:} By following the guidelines for conducting systematic mapping studies in software engineering, we sought to answer research questions on types of venues in which BDD papers have been published, research, contributions, studied topics and their evolution, as well as evaluation methods used in published BDD research. 
 
 \textit{Results:} The study identified 166 papers which were mapped. Key results include the following: the dominance of conference papers; scarcity of research with insights from the industry; shortage of philosophical papers on BDD; acute shortage of metrics for measuring various aspects of BDD specifications and the processes for producing BDD specifications; the dominance of studies on using BDD for facilitating various software development endeavours, improving the BDD process and associated artefacts, and applying BDD in different contexts; scarcity of studies on using BDD alongside other software techniques and technologies; increase in diversity of studied BDD topics; and notable use of case studies and experiments to study different BDD aspects.  
 
 \textit{Conclusion:} The paper improves our understanding of the state of the art of BDD, and highlights important areas of focus for future BDD research.  
 
\end{abstract}

\begin{keyword}
\texttt{Behaviour Driven Development}\sep Systematic Mapping Study\sep Systematic Mapping Studies in Software Engineering 
\end{keyword}

\end{frontmatter}

\nolinenumbers

\section{Introduction}
Behaviour Driven Development (BDD) \cite{dnorth2006, wynne2012} is an agile technique in which software requirements are specified in a semi-structured natural language using Given-When-Then to express examples (also called \textit{scenarios}) of expected software  behaviour or how a user will interact with a software system. Apart from acting as software requirements, the scenarios can also act as test cases, which can be used to verify if the software is behaving as expected. Scenarios are connected to the production code of the System Under Test (SUT) using glue code.  The use of glue code means that the scenarios can be executed to determine the correct and problematic parts of the SUT.

Since BDD was introduced about two decades ago, the Software Engineering research community has investigated various aspects of BDD. Additionally, BDD is becoming an established industry practice for software development and is currently used in different domains~\cite{binamungu2018maintaining}. However, little is known about what has been studied and what has not been studied regarding BDD due to the lack of existing systematic mapping studies as well as systematic literature reviews on the topic. This creates a need to identify what has been most studied and to what extent in order to identify gaps that could inform future research on BDD. A few secondary studies have been conducted about BDD \cite{abushama2020effect, egbreghts2017literature, lillnor2020systematic}. However, these studies were about the impacts of BDD on software projects \cite{abushama2020effect} or were student projects that never got published in peer-reviewed scientific literature \cite{egbreghts2017literature, lillnor2020systematic}. This poses a challenge regarding the availability and reliability of peer-reviewed secondary studies that could inform future research on BDD. Additionally, no peer-reviewed published study has attempted to create the landscape of BDD research. To fill this gap, we followed the guidelines by Petersen \textit{et al.} \cite{petersen2015guidelines, petersen2008systematic} to conduct a Systematic Mapping Study on BDD, to understand the types of venues in which BDD papers have been published, types of research and contributions in published BDD studies, focus and evolution of research on BDD, as well as evaluation methods that have been used in BDD studies.

We found that BDD research has been dominated by conference papers over journal and workshop papers. Also, the majority of BDD studies were found to have focused on three aspects: using BDD for facilitating various software development endeavours, improving the BDD process and associated artefacts, and applying BDD in different contexts. Moreover, there has been an increase in the diversity of studied BDD topics, and there is a notable use of case studies and experiments to evaluate the contributions of different BDD studies. However, there is a scarcity of BDD research that has been evaluated in industry settings. This leads to a lack of understanding of the applicability of the proposed solutions because of a lack of evidence from the software industry. Too, there is a scarcity of studies on using BDD alongside other software development techniques and technologies. Besides, this study has identified a shortage of philosophical papers on BDD, and an acute shortage of metrics for measuring various aspects of BDD, such as metrics for BDD specifications and metrics for processes of producing BDD specifications.

The rest of this paper is structured as follows. Section~\ref{methodology-section} describes the methodology we followed to search, filter and analyse the papers; Section~\ref{results-and-discussion-section} presents the results of the study; Section~\ref{discussion-section} discusses the results and their implications; Section~\ref{future-research-section} presents notable areas of future research based on the results of this study; Section~\ref{validity-threats-section} presents the threats to the validity of our results, and how the validity threats were mitigated; Section~\ref{related-work-section} presents the related work; and Section~\ref{conclusion-section} concludes the paper. 

\section{Research Methodology}
\label{methodology-section}
\begin{figure}[h]
    \centering
    \includegraphics[width=\textwidth]{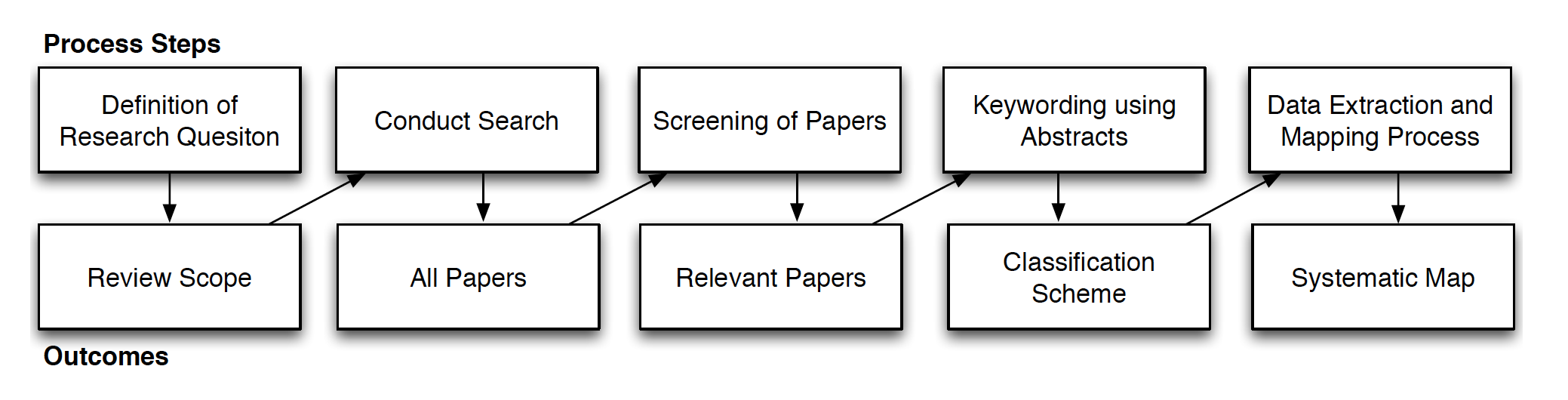}
    \caption{Systematic mapping process ~\cite{petersen2008systematic}}
    \label{fig:systematic_mapping_process}
\end{figure}

To understand the studied BDD topics and provide a structure of the research field, a systematic mapping study is an appropriate research methodology. Systematic mapping studies provide a coarse-grained view of a specific research field by providing visual representations of the research topics as well as the results in the field~\cite{petersen2008systematic}.
This study was conducted by following the steps to conduct a systematic mapping study described by Petersen et al.~\cite{petersen2008systematic}, as shown in Figure~\ref{fig:systematic_mapping_process}. The details of how each step was conducted are provided in the following subsections. 

\subsection{Definition of Research Questions}
\label{subsec:research_questions}
The purpose of the study is to analyse primary research papers on BDD in order to provide an understanding and structuring of different BDD aspects studied in the literature. To derive the research questions, two researchers conducted a brainstorming session, which was aimed at getting a quick overview of existing research on BDD by looking at different papers and referring to existing systematic mapping studies in software engineering. The interest of this systematic mapping is to understand the following aspects with respect to BDD research: timeline, frequency of publications, publication venues, research topics, methodologies used, and achieved results. To this end, the following research questions were formulated: 

\begin{description}
\item[RQ1: ] What are the types of venues in which BDD studies have been published?
\item [RQ2: ] Who is working on BDD research?
\item[RQ3: ] What type of research has been conducted on BDD?
\item[RQ4: ] What types of contributions have been made by existing research on BDD?
\item[RQ5: ] What research evaluation methods have been used in BDD research?
\item[RQ6: ] Which themes have been covered in existing BDD research?
\item[RQ7: ] How have the studied BDD topics evolved over time? 
\item[RQ8: ] How are the contributions in BDD research distributed with respect to research types and topics?
\end{description}

\subsection{Conducting the search}
\label{search-section}
In this step, we searched for scientific papers for use in the study. Five scientific databases were selected due to their relevance in indexing papers published in the field of software engineering and computer science in general (cf.Table~\ref{table:Databases_and_number_of_papers}). Database search is the most commonly used strategy for searching papers in systematic mapping studies in software engineering \cite{petersen2015guidelines}. We also conducted snowballing to identify BDD papers that may not have been obtained through database search. Both forward and backward snowballing were conducted.  
To derive the search string, we scanned titles and abstracts of already known BDD research papers, as well as used the experience of the researchers to come up with keywords that are common in BDD-related research papers. The initial set of keywords was used to derive other keywords in form of synonyms or spellings in British English and American English. The following search string is an example of the search string that was used in the IEEE Xplore database. Since each scientific database uses a slightly different syntax--for instance, the need for brackets and double quotes--the semantics of the search string remained the same but syntactic variations of the search string were made to make it match the syntax of the different databases. 
 
 \texttt{``Behaviour Driven Development” OR ``Behavior Driven Development” OR ``Behavioural software development” OR ``Given when then”}

Since our interest was to survey all papers on BDD, the search string is generic and therefore the number of papers found was also high (791).

\begin{table}[h]
\begin{tabular}{@{}ll@{}}
\toprule
Database            & Search results \\ \midrule
Scopus              & 188            \\
ACM Digital Library & 119            \\
IEEEXplore          & 55             \\
ScienceDirect       & 61             \\
Springer Link       & 368            \\ \midrule
Total & 791 \\ \bottomrule
\end{tabular}
\caption{Number of studies found in each database}
\label{table:Databases_and_number_of_papers}
\end{table}

\subsection{Screening of papers}
The screening of papers was done in three phases. 
In the first phase, we removed all duplicate papers and remained with 628 papers. We then excluded papers that were not written in English. Our search returned some papers which were in Russian and Portuguese and these were removed. This led to a total of 600 papers included. 

In the second phase, we used a tool called Abstrackr~\cite{Abstrackr} to structure and review all the remaining papers. Abstrackr is an online web-based tool that provides the functionality for uploading and organising search results (citation information and abstracts), allowing the reviewers to screen them in a collaborative manner. While Abstrackr also has a feature for automatically screening the abstracts for inclusion or exclusion, we did not use this feature to avoid the probability of introducing errors.

A pilot screening phase using Abstrackr was conducted, where 20 papers were reviewed by both authors, out of which the authors agreed on the judgements made on 14 papers. 
The six conflicts were discussed until a consensus was reached, and this improved our common understanding of how to further judge the papers in subsequent iterations. After the pilot stage and the discussion session, we divided the remaining papers into two, where one researcher reviewed 290 papers and the other 290 papers individually using the following inclusion and exclusion criteria.

Inclusion criteria:
\begin{itemize}
\item The title and abstract show that the paper is about BDD
\item The paper reports a primary study
\item The paper is peer-reviewed
\end{itemize}

Exclusion criteria:
\begin{enumerate}
\item Papers mentioning BDD in the title or abstract but could not be considered as describing research on BDD.
\item Papers presenting editorials and summaries of conferences
\item Papers published in non-peer-reviewed venues
\item Papers not accessible in full text.
\item Papers that are duplicates of other studies
\item General book chapters that explain what BDD is
\item Papers of poor quality that lack sound methodology, clear statement of aims and/or have no contribution.
\end{enumerate}

\begin{figure}[h]
    \centering
    \includegraphics[width=\textwidth]{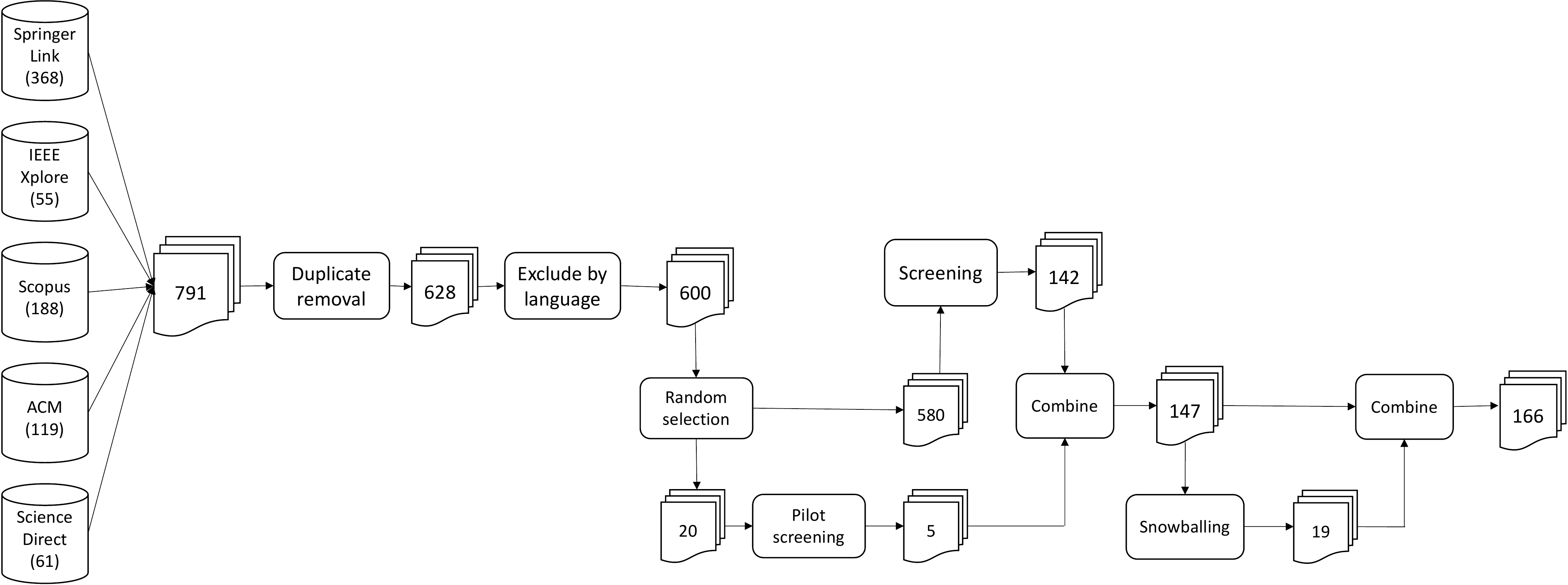}
    \caption{The papers screening process}
    \label{fig:screening_process}
\end{figure}

For some papers, the decision for including or excluding was made based on the abstract, but for others, the abstract was not enough to make a decision, so the reviewers skimmed the full papers to get more information. While working individually, reviewers were allowed to include or exclude a paper if they were completely sure that it meets a certain inclusion or exclusion criteria; in case an individual reviewer was not sure whether to include a paper on not, they marked a paper as ``maybe'' for later discussion and consensus. Thus, after the individual screening, the two reviewers had another meeting to go through all the papers marked as ``maybe" to decide on whether to include or exclude the papers. After this meeting, the final set of included papers was 147. This number also consists of the papers included during the pilot screening. To make sure that we did not miss any valuable papers, we conducted both forward and backward snowballing on all 147 included papers. The snowballing process led to an addition of 19 papers which were not captured by our initial search. Figure~\ref{fig:screening_process} summarises the whole papers screening process.

\subsection{Keywording using Abstracts}
To create the classification scheme of the included papers, four facets (research type, contribution type, research evaluation method and research theme) were selected. We selected these facets because categorising the papers using the facets enables us to answer the research questions posed in Section~\ref{subsec:research_questions}. These facets are explained next.
\begin{description}
\item [Research type]: For each paper, we identified the type of research that the study conducted fits in. We adopted the research type classification by Wieringa et. al.~\cite{wieringa2006requirements}, which is also the classification scheme recommended by Petersen et al.~\cite{petersen2015guidelines}. The classification consists of six types of research which are validation research, evaluation research, solution proposal, philosophical papers, opinion papers and experience papers, as described in Table~\ref{research-types-desc}.

\begin{table*}[htbp]
  \centering
  \caption{Research types~\cite{petersen2015guidelines, wieringa2006requirements}}
\begin{tabular}{p{0.05\linewidth}  p{0.30\linewidth}  p{0.65\linewidth}}
\hline
 {\bf S/n} & {\bf Research Type} & {\bf Description} \\
\hline
         1 & Validation Research & Techniques investigated are novel and have not yet been implemented in practice. Techniques used are for example experiments, i.e., work done in the lab. \\
\hline
         2 & Evaluation Research & Techniques are implemented in practice and an evaluation of the technique is conducted. That means, it is shown how the technique is implemented in practice (solution implementation) and the consequences of the implementation in terms of benefits and drawbacks (implementation evaluation). This also includes identifying problems in the industry. \\
\hline
         3 & Solution Proposal & A solution for a problem is proposed, the solution can be either novel or a significant extension of an existing technique. The potential benefits and the applicability of the solution are shown by a small example or a good line of argumentation. \\
\hline
         4 & Philosophical Papers & These papers sketch a new way of looking at existing things by structuring the field in the form of a taxonomy or conceptual framework. \\
\hline
         5 & Opinion Papers & These papers express the personal opinion of somebody on whether a certain technique is good or bad, or how things should be done. They do not rely on related work and research methodologies. \\
\hline
         6 & Experience & Papers explain what has been done in practice, and how it has been done. It has to be the personal experience of the author. \\ 
\hline
\end{tabular}
\label{research-types-desc}%
\end{table*}%

\item [Contribution type]: Similarly, for each of the included papers, we identified the contribution type of the reported study. In particular, we adopted the following five types of contributions from Mujtaba \cite{mujtaba2008software}: process, method, model, tool, and metric. The five contribution types are described in Table~\ref{contribution-types-desc}. We devised a sixth contribution type called \textit{Empirical Insights} to categorize studies whose contributions do not fall into any of the above five contribution types, but still report important empirical insights regarding BDD. 

\begin{table*}[htbp]
  \centering
  \caption{Contribution types~\cite{mujtaba2008software}}
\begin{tabular}{p{0.05\linewidth}  p{0.30\linewidth}  p{0.65\linewidth}}
\hline
 {\bf S/n} & {\bf Contribution Type} & {\bf Description} \\
\hline
         1 &    Process & Describes actions or activities, their associated workflows and artifacts\\
\hline
         2 &     Method & Describes rules for doing things. It  algorithms \\
\hline
         3 &      Model & Describes the real world without giving details. It should be very formal with semantics and notations, e.g., UML models \\
\hline
         4 &       Tool & A software tool has been developed to support different BDD aspects \\
\hline
         5 &     Metric & Describes metrics and measurements for different aspects of a BDD process and the resulting specifications \\
\hline

         6 &     Empirical insights & Describes lessons or experience of using BDD in practice. The lessons are not based on the evaluation of a process, model, method, tool, or metric. For example, a study that mainly reports the benefits and challenges of using BDD in a particular context \cite{nascimento2020behavior} would be regarded to have contributed empirical insights. \\
\hline
\end{tabular}
\label{contribution-types-desc}%
\end{table*}%

\item [Research evaluation method]: For each paper, we analysed how the contribution of the paper was evaluated. Specifically, we used the research evaluation methods classification scheme proposed by Chen and Babar~\cite{chen2011systematic} in their systematic literature review (refer to Table~\ref{evaluation-method-types-desc}).

\begin{table*}[htbp]
  \centering
  \caption{Types of evaluation methods~\cite{chen2011systematic}}
\begin{tabular}{p{0.05\linewidth}  p{0.30\linewidth}  p{0.65\linewidth}}
\hline
 {\bf S/n} & {\bf Evaluation Method} & {\bf Description} \\
\hline
         1 & Rigorous Analysis & Involves thorough analysis, deriving, or proving certain aspects of a system, e.g., in formal models \\
\hline
         2 & Case Study & Uses multiple sources of evidence to investigate a phenomenon in its real world context, when there are unclear boundaries between a phenomenon and its context. 
         \\
\hline
         3 & Discussion & Evaluation is based on qualitative, textual, opinions, to compare and contrast some aspects of a proposed solution or discuss the upsides and downsides of a proposed solution \\
\hline
         4 &    Example & An example is used to both explain what has been proposed by authors and demonstrate how good a proposal is \\
\hline
         5 & Experience Report & Using evidence collected formally or informally to demonstrate the use of a particular proposal, without using case studies or controlled experiments \\
\hline
         6 & Experiment & Testing an hypothesis through controlled experiments in industry or laboratory settings \\
\hline
         7 & Simulation & Artificial data have been used to execute a system, by modelling the real world \\
\hline
\end{tabular} 
\label{evaluation-method-types-desc}%
\end{table*}%

\item[Research theme:] The research themes were derived in an inductive manner. The two researchers read the abstract and screened the papers (where necessary) to identify thematic areas covered by the paper. Note that one paper can cover more than one theme. 
\end{description}

To place the papers into specific categories, the two researchers had three workshops where 50 randomly selected included papers were discussed and the research type, contribution type, and research theme were decided. This was to make sure that both the researchers agree to the categories, especially in the research theme facet, since the categories were decided on in an inductive manner rather than a deductive manner. Categorising 50 papers together also meant that the researchers have a common understanding of the categories used. The remaining set of papers (97) was divided almost equally among the two researchers, and each researcher continued with the categorisation independently. After the categorisation of papers by individual researchers, a meeting of researchers was held to discuss the categorisation results and to resolve any challenges encountered by individual researchers when assigning themes to papers. 

\subsection{Data Extraction and Mapping process}
The categorisation of each paper was done in an Excel file where for each paper we marked the following: research type, research theme, contribution type, and evaluation method from the key-wording stage. Additionally, for each paper, we marked the year of publication, all authors, as well as the publication type (journal, book chapter, conference or workshop). The frequencies of the publications in each of these categories were calculated with the help of Excel formulas.

\section{Results}
\label{results-and-discussion-section}
We now present the results and discuss them in Section~\ref{discussion-section}.

\subsection{RQ1: Types of venues that have published BDD papers}
\begin{figure}[!htb]
\centering
  \includegraphics[scale=0.38]{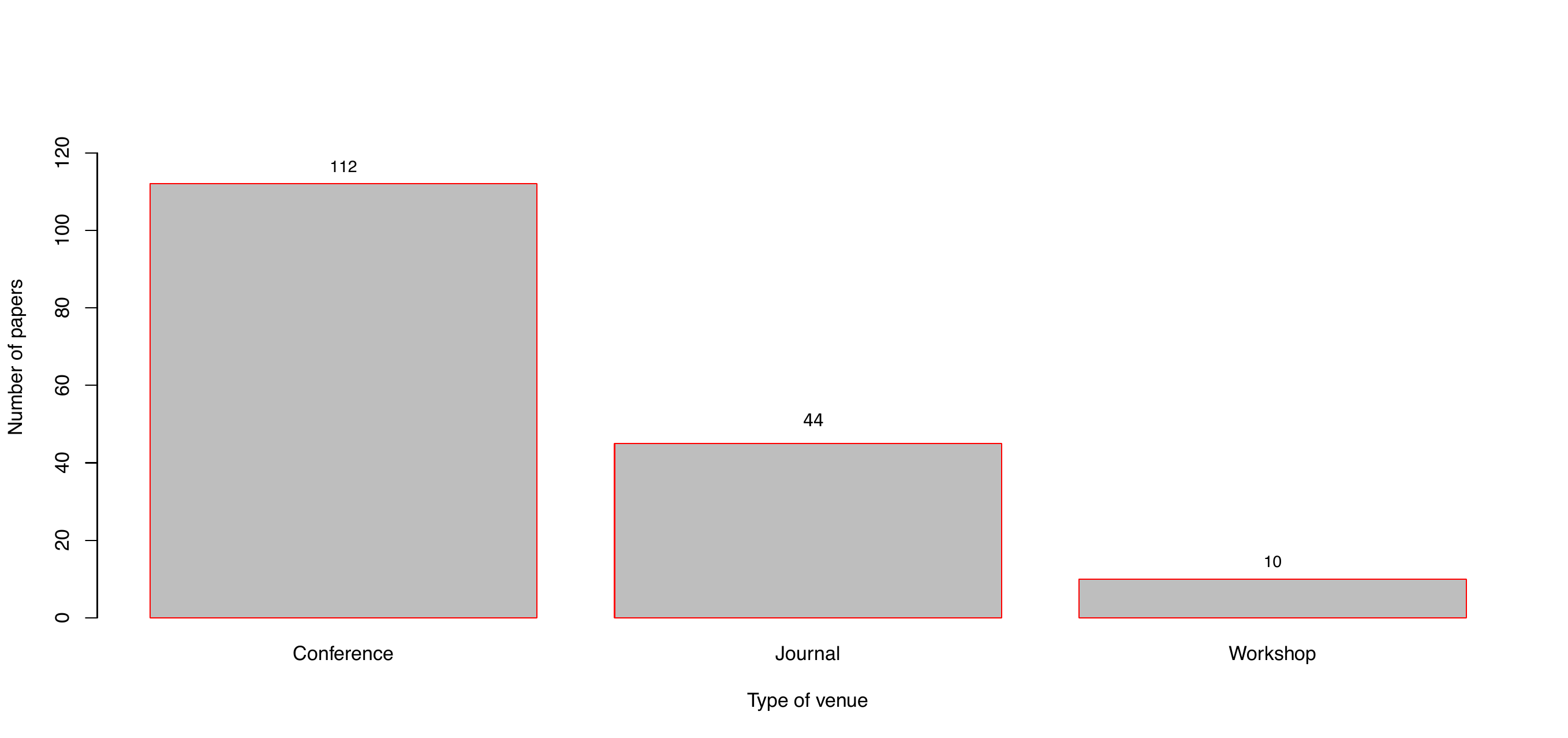}
   \vspace*{-3mm}
 \caption{Distribution of publications by venue}
 \label{distribution-of-publications-by-venue}
\end{figure}

Figure~\ref{distribution-of-publications-by-venue} shows the number of publications in different venues. It is observed that most papers were published in conferences, followed by journals and workshops. Further analysis of the conference papers shows that the papers were published in a variety of conferences. As seen in Table\ref{table_conference_distribution}, the study found that the conference papers were published in 89 different conferences. This shows that apart from the International Conference on Agile Software Development (formally called XP) which has 7 papers and the Brazilian Symposium on Software Engineering which has 5 papers, the rest of the conferences have less than 5 papers on BDD over the years (cf. Table~\ref{table_conference_distribution}). A similar trend is observed in the journal publications, the 44 journal papers are scattered in 38 different journals. Out of the 38 journals, Information and Software Technology (IST) has three papers and the Journal of Systems and Software (JSS) has three papers, while the rest of the journals have one paper each (cf. Table~\ref{table_journal_distribution}). Additionally, all workshop papers are also published in distinct workshops.

\begin{table}[]
\begin{tabular}{@{}p{0.9\textwidth}p{0.1\textwidth}@{}}
\toprule
Conference                                                                           & No. of papers \\ \midrule
Agile Software Development (XP)                                                      & 7                \\
Brazilian Symposium on Software Engineering                                          & 5                \\
Model Driven Engineering Languages and Systems (MODELS)                              & 4                \\
Iberian Conference on Information Systems and Technologies                           & 3                \\
Computational Science and Its Applications                                           & 2                \\
Conceptual Modelling                                                                  & 2                \\
Information Technology - New Generations                                             & 2                \\
Requirements Engineering                                                             & 2                \\
Requirements Engineering: Foundation for Software Quality                            & 2                \\
Brazilian Symposium on Systematic and Automated Software Testing                     & 2                \\
Symposium on Engineering Interactive Computing Systems                               & 2                \\
International Conference on the Quality of Information and Communications Technology & 2                \\
Other conferences                                                                               & 77               \\ \bottomrule
\end{tabular}
\caption{Distribution of papers by conferences.}
\label{table_conference_distribution}
\end{table}

\begin{table}[]
\begin{tabular}{@{}ll@{}}
\toprule
Journal                             & No. of papers \\ \midrule
Information and Software Technology & 3             \\
Journal of Systems and Software     & 3             \\
Other journals                      & 38           
\end{tabular}
\caption{Distribution of papers by journals.}
\label{table_journal_distribution}
\end{table}
\subsection{RQ2: Researchers who are working on BDD}
\begin{figure}[!htb]
\centering
  \includegraphics[scale=0.8]{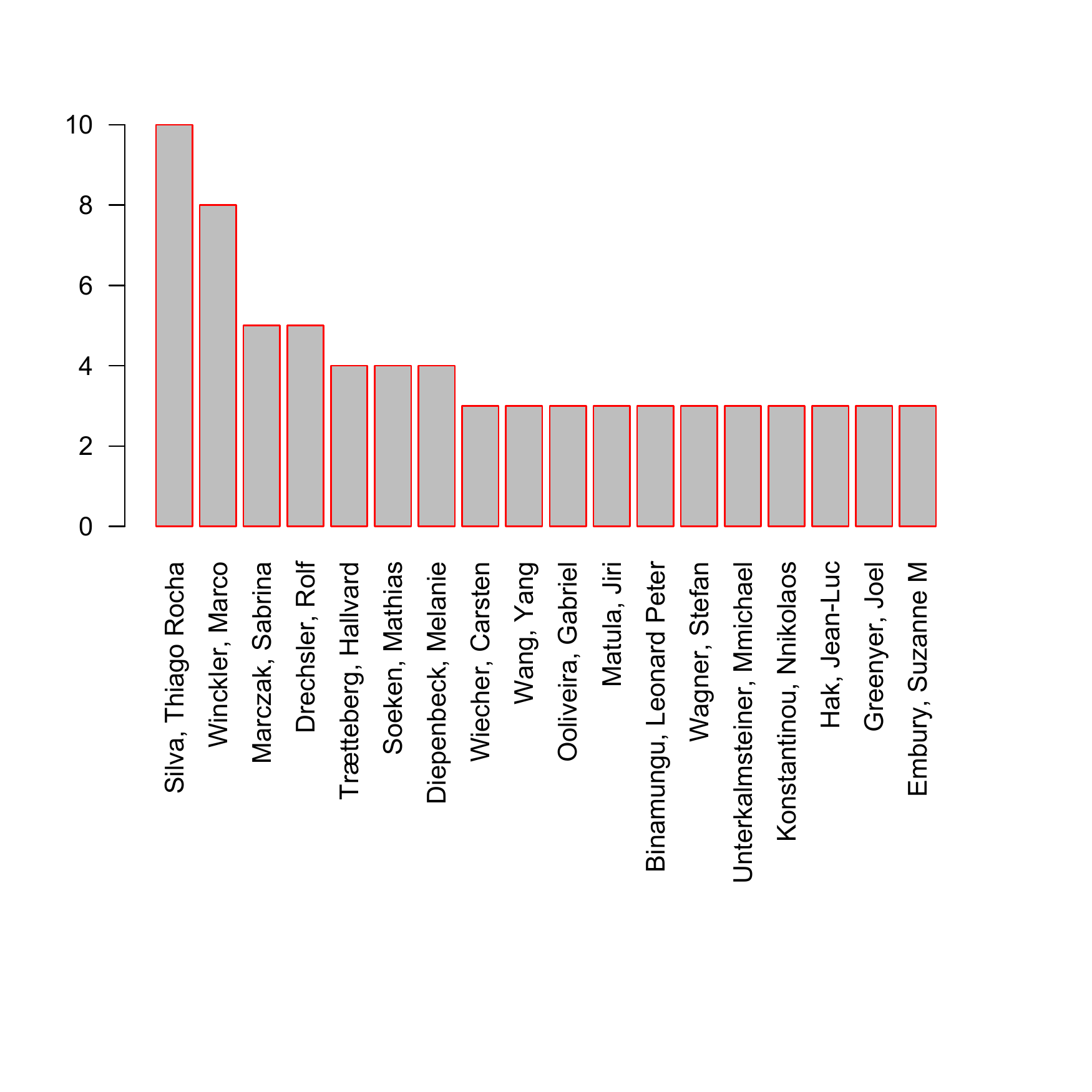}
   \vspace*{-3mm}
 \caption{Top authors in the BDD research area.}
 \label{top-authors}
\end{figure}

We analysed all 166 papers to identify the authors, with the aim of identifying top researchers in the BDD area. There are a total of 640 unique authors and co-authors. However, further analysis shows that most authors only have one or two publications in the field. The top ten authors in terms of publication
frequency are shown in Figure~\ref{top-authors}, where the author with most publications has
published 10 of the included papers and there are 11 authors in the 8th position
who have published 3 papers. Note that this analysis includes both first authors and co-authors. 

\subsection{RQ3: Types of conducted BDD research}
\label{conducted-research-section}
\begin{figure}[!htb]
\centering
  \includegraphics[scale=0.38]{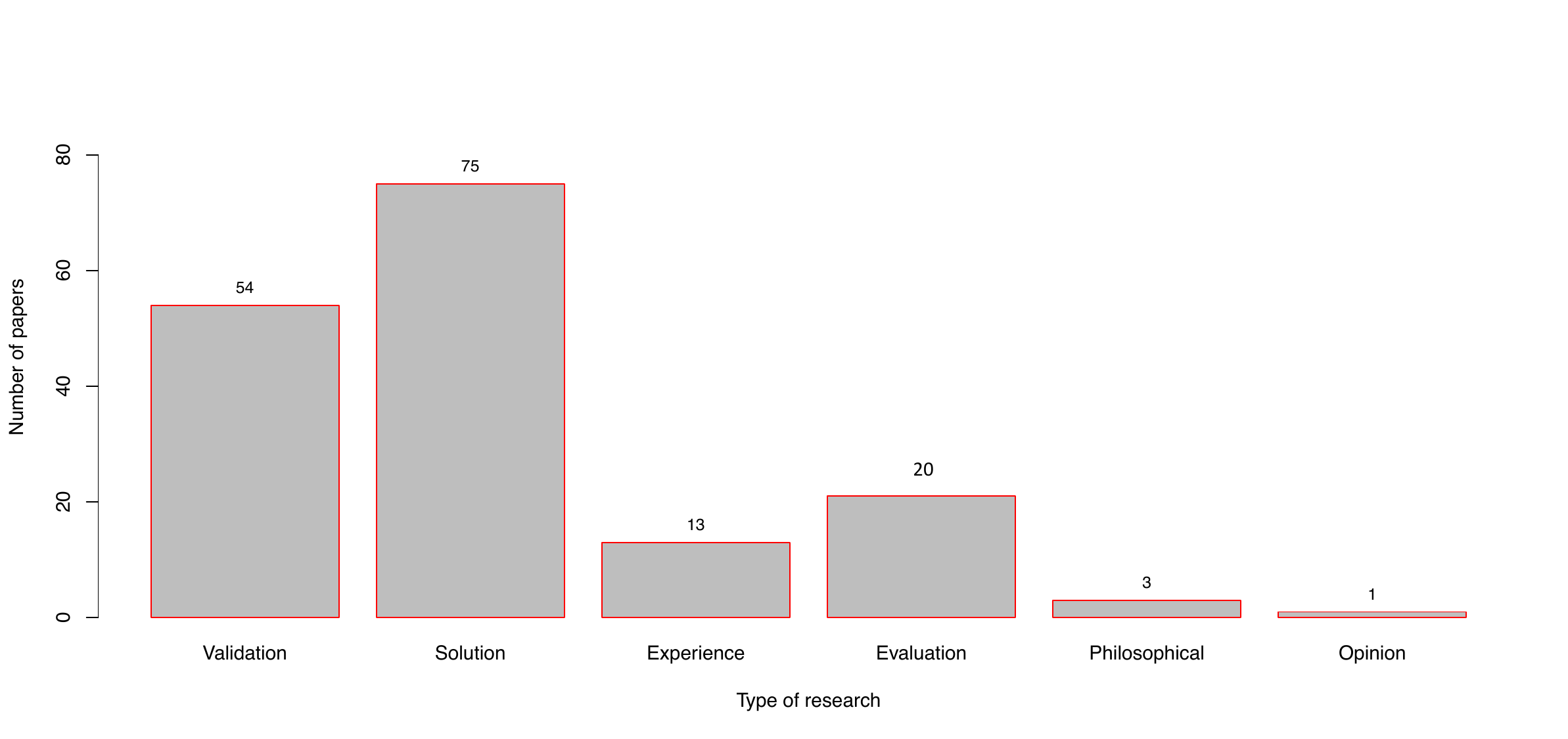}
   \vspace*{-3mm}
 \caption{Distribution of types of conducted BDD research}
 \label{distribution-of-research-types}
\end{figure}

Figure~\ref{distribution-of-research-types} shows the distribution of different research types conducted about BDD. On the one hand, there are more solution proposals and validation research than evaluation research. On the other hand, there are a few papers reporting the experience of using BDD in practice. Also, philosophical and opinion papers are almost non-existent. Table~\ref{research-types-studies} summarises the studies under each research type.

\subsection{RQ4: Types of contributions in conducted BDD research}
\begin{figure}[!htb]
\centering
  \includegraphics[scale=0.38]{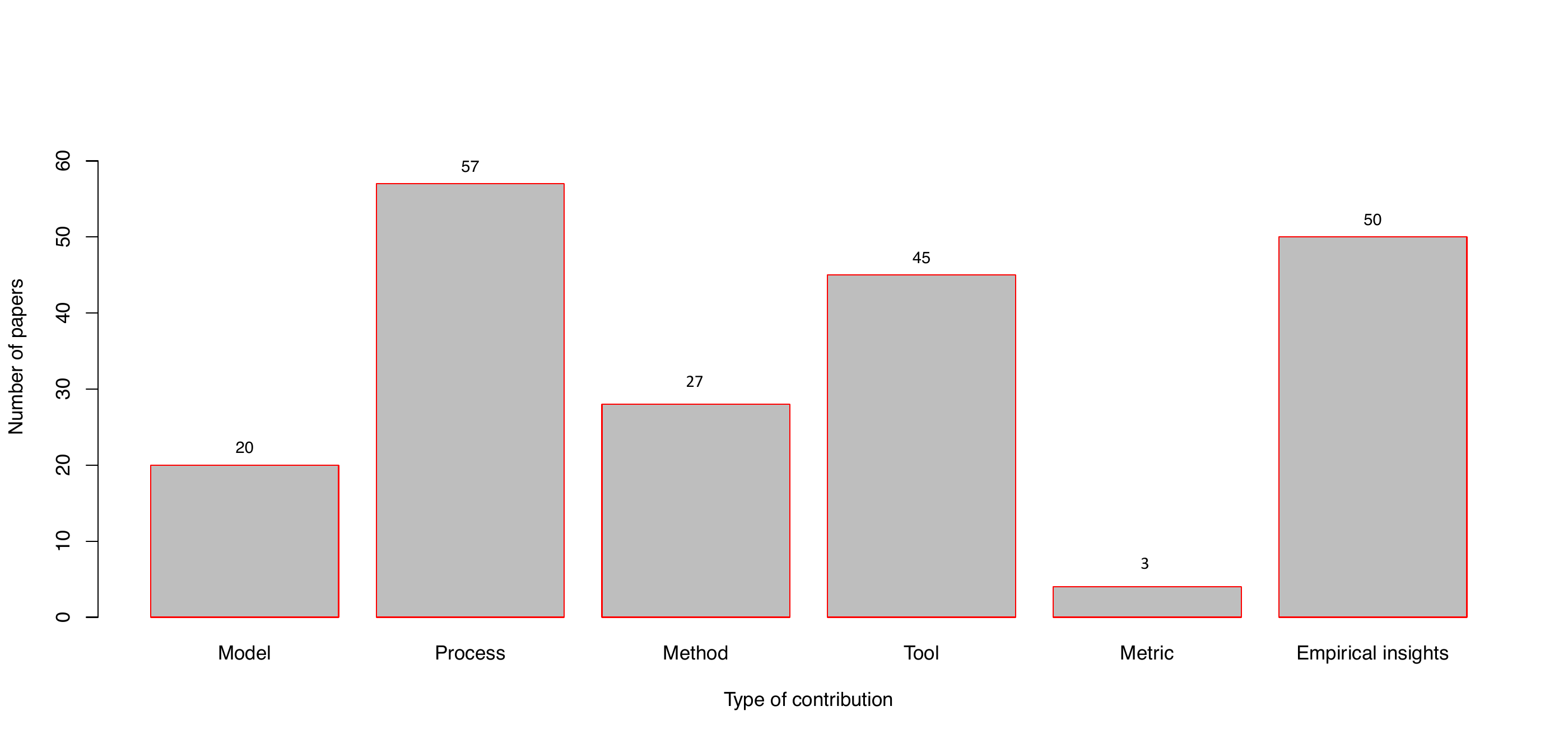}
   \vspace*{-3mm}
 \caption{Distribution of types of contributions in conducted BDD research}
 \label{distribution-of-contribution-types}
\end{figure}

Figure~\ref{distribution-of-contribution-types} shows the distribution of different contributions made by studies on BDD. In general, BDD studies have proposed more processes, tools, and empirical insights than methods, models, and metrics. Table~\ref{contribution-types-studies} summarises the studies under each contribution type.

\subsection{RQ5: Research evaluation methods used}
\label{research-methods-section}
\begin{figure}[!htb]
\centering
  \includegraphics[width=\textwidth]{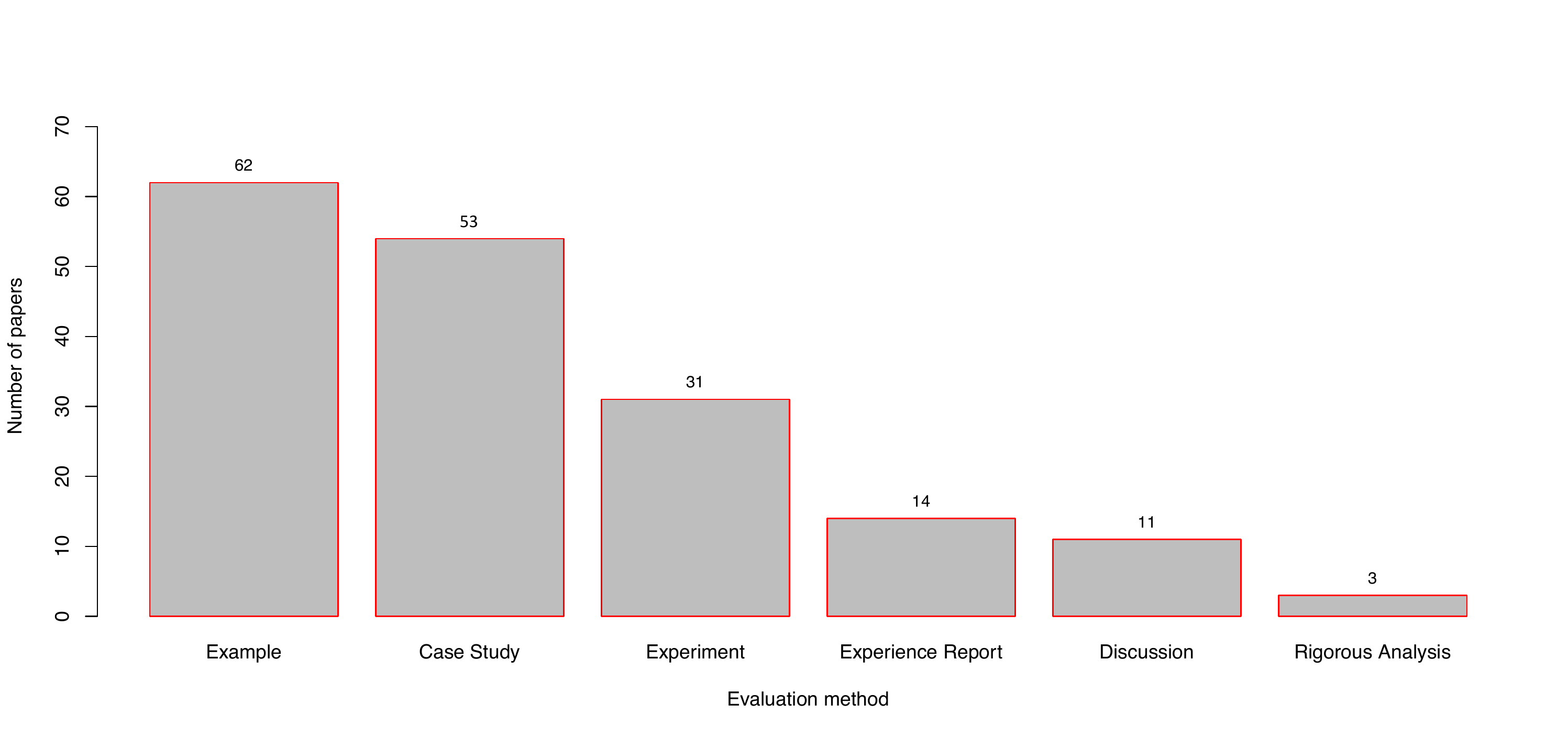}
   \vspace*{-3mm}
 \caption{Evaluation methods used in BDD research}
 \label{research-methods}
\end{figure}

Figure~\ref{research-methods} shows the different research evaluation methods that have been used in BDD research. Because some papers have used more than one evaluation method, the total in Figure~\ref{research-methods} is more than the total number of papers analysed in the present study. Also, it can be seen in Figure~\ref{research-methods} that most studies have used example applications or case studies to demonstrate the applicability of solutions or insights generated in BDD research. Table~\ref{evaluation-methods-studies} summarises the studies under each evaluation method.  

\subsection{RQ6: Themes in existing BDD research}
\label{themes-results-section}
\begin{figure}[!htb]
\centering
  \includegraphics[width=\textwidth]{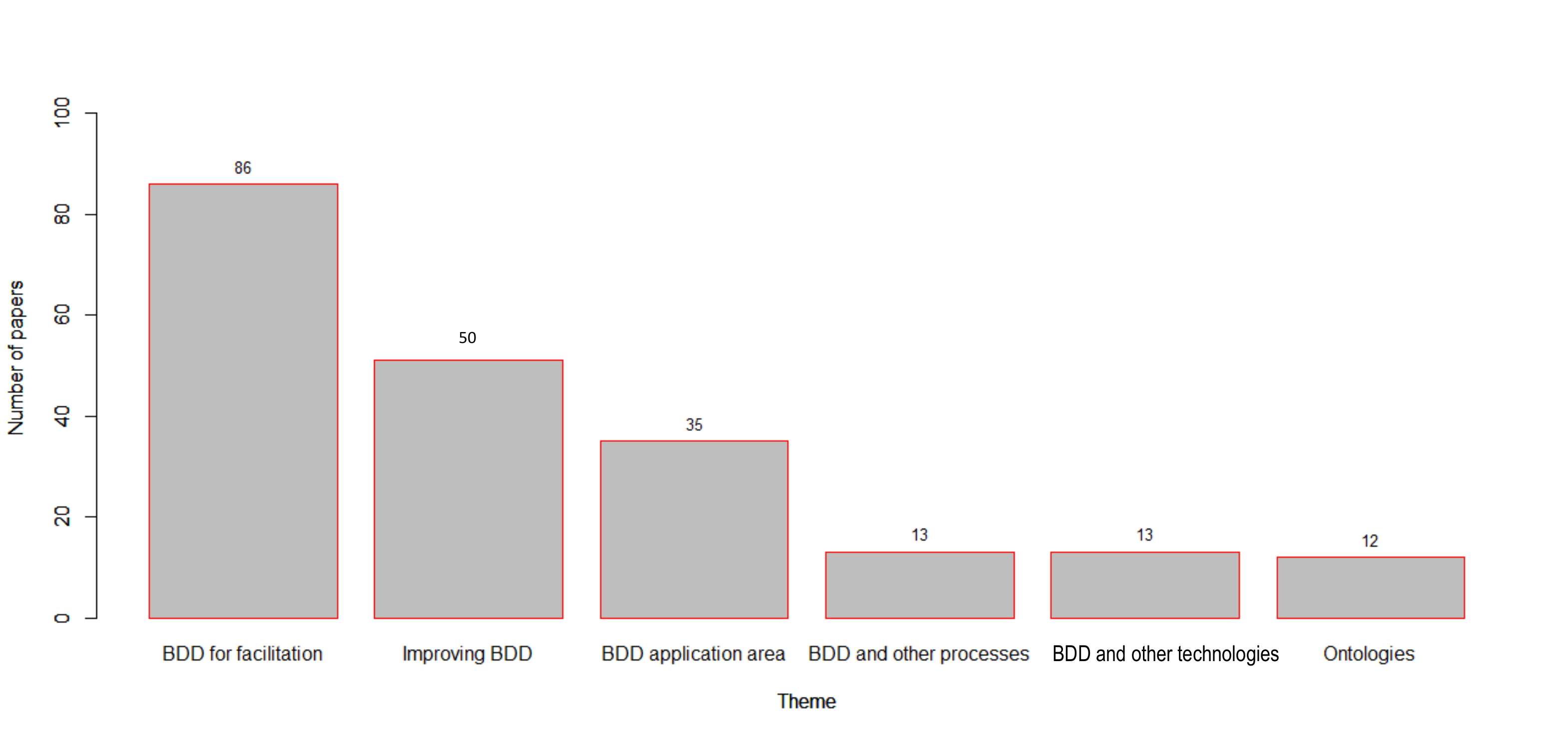}
   \vspace*{-3mm}
 \caption{Distribution of themes in BDD research}
 \label{themes-distribution}
\end{figure}

Our systematic mapping identified six major themes, which are presented in Figure~\ref{themes-distribution} and described in this subsection. Each major theme had several sub-themes which are more specific, and the papers are therefore categorised using sub-themes. The themes and sub-themes are summarised in Table \ref{themes-and-sub-themes}, while the studies under each theme are presented in Table ~\ref{existing-themes}.
It can be seen that BDD research has mainly focused on how BDD is used to facilitate various software engineering aspects, improving the BDD process and the resulting artefacts, and on how BDD has been used in different application areas. Studies on using ontologies, BDD alongside other processes and BDD with other technologies are still on the lower side. 

\begin{longtable}{p{0.02\linewidth}  p{0.2\linewidth}  p{0.40\linewidth} }
\caption{Themes in existing BDD research}
    \label{themes-and-sub-themes}\\
\hline
 {\bf Sn} & {\bf Theme} & {\bf Sub-theme(s)} \\
\hline
\multicolumn{ 1}{r}{1} & \multicolumn{ 1}{l}{Facilitating software development} &     Safety  \\
\multicolumn{ 1}{c}{} & \multicolumn{ 1}{c}{} & GUI testing  \\
\multicolumn{ 1}{c}{} & \multicolumn{ 1}{c}{} & Artifact generation \\
\multicolumn{ 1}{c}{} & \multicolumn{ 1}{c}{} & Product lines \\
\multicolumn{ 1}{c}{} & \multicolumn{ 1}{c}{} & Compliance testing \\
\multicolumn{ 1}{c}{} & \multicolumn{ 1}{c}{} & Traceability \\
\multicolumn{ 1}{c}{} & \multicolumn{ 1}{c}{} & Model testing  \\
\multicolumn{ 1}{c}{} & \multicolumn{ 1}{c}{} & Business process modelling \\
\multicolumn{ 1}{c}{} & \multicolumn{ 1}{c}{} & Security and privacy  \\
\multicolumn{ 1}{c}{} & \multicolumn{ 1}{c}{} & Model-based testing \\
\multicolumn{ 1}{c}{} & \multicolumn{ 1}{c}{} & Integration testing \\
\multicolumn{ 1}{c}{} & \multicolumn{ 1}{c}{} & Load testing  \\
\multicolumn{ 1}{c}{} & \multicolumn{ 1}{c}{} & Syntax highlighting \\
\multicolumn{ 1}{c}{} & \multicolumn{ 1}{c}{} & Reproducing bugs  \\
\multicolumn{ 1}{c}{} & \multicolumn{ 1}{c}{} & Spreadsheets testing \\
\multicolumn{ 1}{c}{} & \multicolumn{ 1}{c}{} & Test selection   \\
\multicolumn{ 1}{c}{} & \multicolumn{ 1}{c}{} & Usability testing  \\
\multicolumn{ 1}{c}{} & \multicolumn{ 1}{c}{} & Software documentation  \\
\hline
\multicolumn{ 1}{c}{2} & \multicolumn{ 1}{l}{Improving BDD} & Specification quality  \\
\multicolumn{ 1}{c}{} & \multicolumn{ 1}{c}{} & BDD adoption \\
\multicolumn{ 1}{c}{} & \multicolumn{ 1}{c}{} & Co-evolution of artifacts  \\
\multicolumn{ 1}{c}{} & \multicolumn{ 1}{c}{} & Duplicate detection and refactoring \\
\multicolumn{ 1}{c}{} & \multicolumn{ 1}{c}{} & BDD characteristics \\
\multicolumn{ 1}{c}{} & \multicolumn{ 1}{c}{} & Specification language  \\
\multicolumn{ 1}{c}{} & \multicolumn{ 1}{c}{} & Test cases as requirements \\
\multicolumn{ 1}{c}{} & \multicolumn{ 1}{c}{} & BDD tools \\
\hline
\multicolumn{ 1}{c}{3} & \multicolumn{ 1}{l}{BDD application areas} & Hardware context  \\
\multicolumn{ 1}{c}{} & \multicolumn{ 1}{c}{} & Health context \\
\multicolumn{ 1}{c}{} & \multicolumn{ 1}{c}{} & Education context \\
\multicolumn{ 1}{c}{} & \multicolumn{ 1}{c}{} & Mobile apps  \\
\multicolumn{ 1}{c}{} & \multicolumn{ 1}{c}{} & Enterprise systems \\
\multicolumn{ 1}{c}{} & \multicolumn{ 1}{c}{} & BDD and computer networking  \\
\multicolumn{ 1}{c}{} & \multicolumn{ 1}{c}{} & Product configuration systems \\
\multicolumn{ 1}{c}{} & \multicolumn{ 1}{c}{} & Multi-agent systems  \\
\hline
         4 & Ontologies & Ontologies  \\
\hline
\multicolumn{ 1}{c}{} & \multicolumn{ 1}{l}{BDD and other processes} & Comparison with other techniques  \\
\multicolumn{ 1}{c}{} & \multicolumn{ 1}{c}{} & BDD and Scrum \\
\multicolumn{ 1}{c}{} & \multicolumn{ 1}{c}{} & BDD and DevOps  \\
\hline
\multicolumn{ 1}{c}{6} & \multicolumn{ 1}{l}{BDD and other technologies} & Machine Learning  \\
\multicolumn{ 1}{c}{} & \multicolumn{ 1}{c}{} & Blockchain \\
\multicolumn{ 1}{c}{} & \multicolumn{ 1}{c}{} & Microservices \\
           & \multicolumn{ 1}{c}{} & Web services \\
\hline
\end{longtable}

\subsubsection*{Theme 1: Facilitating software development}
In most studies, BDD has been used to facilitate different aspects of a software or software engineering process. In particular, BDD has been used to facilitate the following: safety of software systems \cite{wang2018combiningicse, wang2018combining, baillon2010executable, wiecher2020scenarios, wang2018speed}, testing of graphical user interfaces (GUI) \cite{silva2017formal, bunder2019model, sivanandan2014agile, schur2017augmented, bahaweres2020behavior, rocha2019ensuring, rocha2019ensuringnew, bunder2019towards}, testing of spreadsheets \cite{almeida2016ss}, product line engineering \cite{elshandidy2019behaviour, bagheri2013light, elshandidy2021using}, compliance testing \cite{lopez2014behaviour, moult2020compliance, williams2020legislation, morrison2013proposing, morrison2013proposingnew, pang2021towards}, traceability of software artefacts \cite{lucassen2017behavior, silva2021parsing, yang2019predicting}, model testing \cite{mens2019method, snook2018behaviour, snook2021domain, silva2021empirical, rocha2019ensuring, silva2019extending, fischer2019formal}, business process modelling \cite{de2012business, carvalho2013implementing, lubke2016modeling, matula2018reengineering}, test selection \cite{xu2021requirement}, usability testing \cite{guncan2021user}, security and privacy of software systems \cite{lai2014combining, purkayastha2020continuous, okubo2014security}, model-based testing \cite{sivanandan2014agile, li2016skyfire}, automatic generation of different software artefacts, including various artefacts of a BDD project \cite{wanderley2012framework, bunder2019model, dimanidis2018natural, soeken2012assisted, schur2017augmented, nguyen2020automated, kamalakar2013automatically, malik2021automating, carrera2014beast, storer2019behave, nezhad2018behavior, lazuar2010behaviour, snook2018behaviour, carter2016bhive, carter2016bhive, deng2021bmt, alferez2019bridging, gupta2019creation, pandit2016distributed, silva2021empirical, fazzolino2019feature, gao2016generating, bonfanti2018generation, schoeneman2013integrating, king2014legend, williams2020legislation, nguyen2020automated, li2016skyfire, mahalakshmi2017theoretical, o2018toward, diepenbeck2013towards, bunder2019towards, pinto2019user, siqueira2017using}, integration testing \cite{purkayastha2020continuous, bussenot2018orchestration}, load testing \cite{schulz2019behavior}, syntax highlighting \cite{matula2018ontological}, and reproducing bugs \cite{karagoz2017reproducing}.
 
\subsubsection*{Theme 2: Improving the BDD process and the resulting artefacts}
Another substantial strand of research on BDD has been focusing on improving various aspects of the BDD process and the resulting artefacts. Specifically, studies on improving BDD have covered the following aspects: assessing and improving the quality of BDD specifications \cite{binamungu2020characterising, oliveira2019evaluate, binamungu2018maintaining, oliveira2017empirical, oliveira2018understanding, binamungu2018detecting}, ensuring co-evolution of BDD artifacts \cite{sathawornwichit2012consistency, zampetti2020demystifying, santos2019improving, yang2019predicting, rocha2019using}, adoption of BDD by software teams \cite{irshad2021adapting, contan2017automated, bezsmertnyi2020behavior, gil2016behavior, scandaroli2019behavior, pereira2018behavior, shafiee2018behavior, nascimento2020behavior, zampetti2020demystifying, binamungu2018maintaining, mello2018applicability, cavalcante2019behavior, barus2019implementation}, characteristics of BDD \cite{solis2011study}, duplicate detection and refactoring of BDD specifications \cite{ binamungu2018detecting}, developing a software specification language \cite{kudo2019conceptual, lucena2017semantic, lazuar2010behaviour, deng2021bmt, snook2021domain, rocha2020evaluating, li2017gherkin, haser2016business, king2014legend, hesenius2014towards, pang2021towards, bunder2019towards}, using test cases as requirements \cite{bjarnason2016multi, bjarnason2015industrial, elshandidy2021using}, and BDD tools \cite{okolnychyi2016study, pyshkin2012requirements}.

\subsubsection*{Theme 3: Application of BDD in different settings}
Different studies have also reported the application of BDD in different areas/contexts. Existing literature shows that BDD has been particularly used in the following settings: development of multi-agent systems \cite{carrera2014beast}, development and testing of systems that have hardware components (hardware context) \cite{kannengiesser2020behaviour, alhaj2019approach, zaeske2021behavior, diepenbeck2012behavior, nezhad2018behavior, diepenbeck2018behaviour, diepenbeck2014behaviour, deng2021bmt, mwakyanjala2020concurrent, winkler2019efficient, bussenot2018orchestration, zafar2021towards, alhaj2017using, cauchi2016using}, use of BDD to develop software systems for use in the health sector (health context) \cite{hatko2014behaviour, purkayastha2020continuous, mello2018applicability, morrison2013proposing, morrison2013proposingnew, giorgi2019transition}, use of BDD to teach software engineering or teaching BDD in software engineering classes (education context) \cite{Sarinho2019essemble, nascimento2020behavior, rocha2021enhancing}, facilitating computer networking \cite{esposito2018behavior}, developing product configuration systems \cite{shafiee2018behavior}, development and testing of mobile applications \cite{nguyen2020automated, ali2019using}, and development of enterprise systems \cite{fazzolino2018assessing, gohil2011towards}.

\subsubsection*{Theme 4: Ontologies}
Another theme that has been identified is on BDD in combination with ontologies. Specifically, this has been studied in two aspects; first, developing an ontology that formalises BDD vocabularies that are used in user stories, scenarios and GUI interactions~\cite{silva2017behavior}. The ontology can be used to build model-driven tools that enable the modelling of BDD scenarios. The existence of such a model also makes it possible to evaluate the consistency of scenarios and other GUI-related artefacts such as GUI prototypes and task models ~\cite{silva2017formal, rocha2019ensuringnew, rocha2020ensuring, rocha2019ensuring, silva2016testing}.  Second, BDD has been used in combination with 
ontologies of different domains with the aim of reducing the ambiguity of defining BDD scenarios using natural language. Our study found BDD used with an ontology in the higher education domain~\cite{souza2018improving, lopes2021scrumontobdd} and an ontology in the business information systems domain~\cite{matula2018enterprise, matula2018ontological}. Similarly, the presence of domain ontologies used in BDD scenarios facilitates consistency checking between the functionality of the system and business-related tasks~\cite{matula2018reengineering}. 

\subsubsection*{Theme 5: BDD and other processes}
BDD has also been used alongside other software engineering processes. Specifically, BDD has been compared with other software development techniques \cite{santos2015study, oran2017analysing, dookhun2019assessing, santos2018automated, bezsmertnyi2020behavior, manuaba2019combination} and combined with other software development methods such as Scrum \cite{de2017combining, souza2018improving, lopes2021scrumontobdd} and DevOps \cite{zaeske2020leveraging, giorgi2019transition}, to generate lessons for informing better software development.

\subsubsection*{Theme 6: BDD and other technologies}
BDD has been studied by integrating it with other technologies. Precisely, lessons have been generated when BDD was used alongside machine learning technology from two perspectives. First, how machine learning can enhance the use of BDD artefacts, for instance by recommending micro-services from a service catalogue based on scenarios selected ~\cite{ma2018scenario}. Second, how BDD can facilitate the development of machine learning applications, for instance by using BDD scenarios to specify accountability requirements for machine learning algorithms~\cite{pang2021towards}. 
A few studies have also been identified with respect to microservices technology where BDD scenarios have been used to select which microservices are relevant ~\cite{ma2018scenario,ma2019graph}, automate acceptance testing of microservices by executing appropriate scenarios~\cite{rahman2015reusable} and using BDD scenarios to solicit crowdsourced microservices~\cite{aghayi2021crowdsourced}. Two studies have also been found where BDD is used to facilitate the development and testing of web services~\cite{dimanidis2018natural,orucc2016testing}. One unique study discussed the use of BDD to support the development of blockchain applications~\cite{liao2017toward}.  

\subsection{RQ7: Evolution of topics over time }
We start by showing the frequency of BDD papers published over the years, and, thereafter, we show how the themes in those papers have evolved over time.

\subsubsection{Publication frequency over the years}

\begin{figure}[!htb]
\centering
  \includegraphics[scale=0.38]{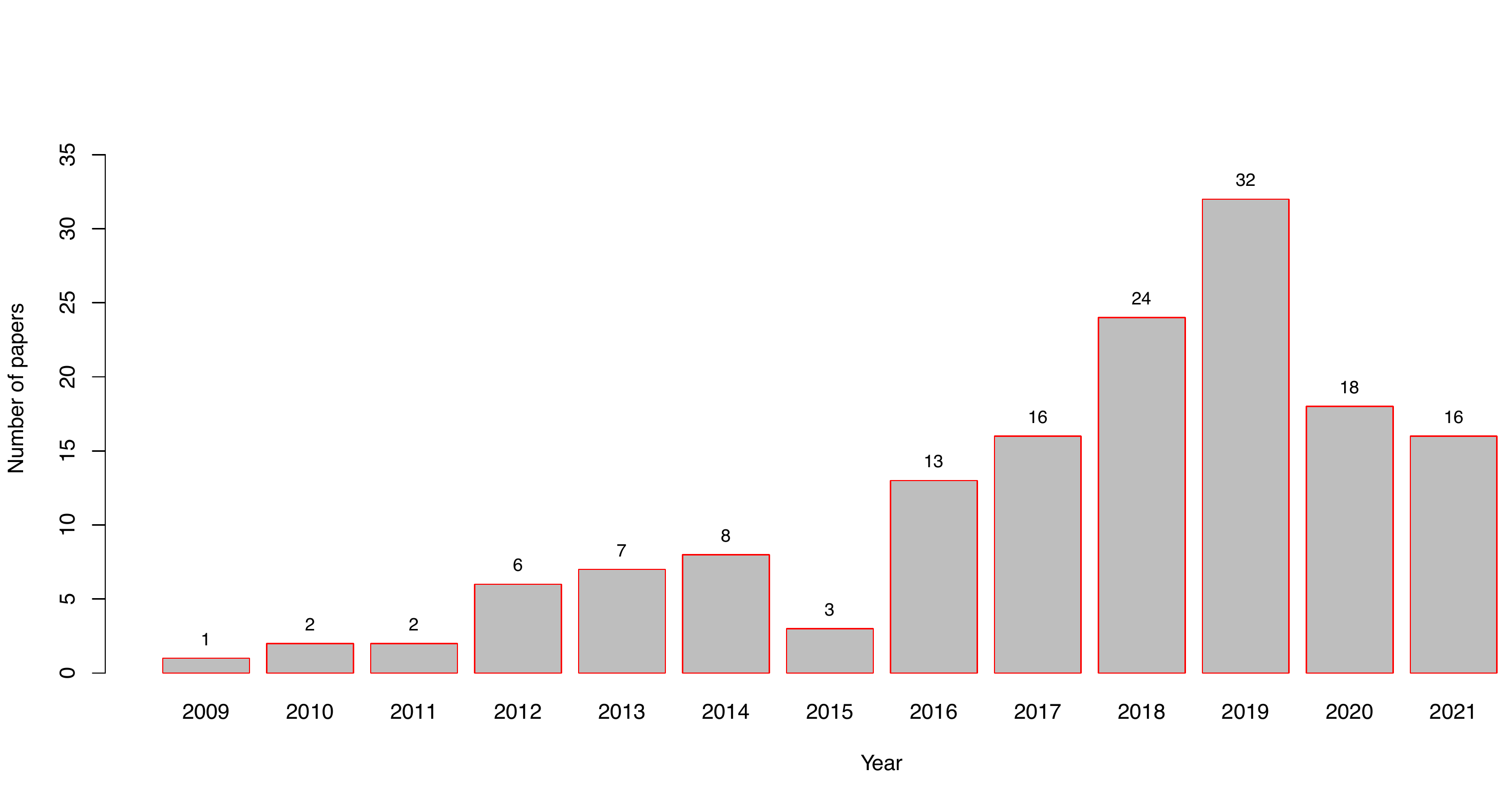}
   \vspace*{-3mm}
 \caption{Frequency of publication over the years (2009 -- 2021)}
 \label{frequency-of-publications-per-year}
\end{figure}

Figure~\ref{frequency-of-publications-per-year} shows the frequency of publication over the years. These results show that, after the introduction of BDD through a magazine article in 2006 \cite{dnorth2006}, there has been an increase in the number of publications on BDD; a peak was reached in 2019, when 32 papers were published. In particular, a notable increase in the number of BDD papers was observed from 2016 to 2021. 
 
\subsubsection{Evolution of studied BDD topics over the years}
\begin{figure}[!htb]
\centering
  \includegraphics[width=\textwidth]{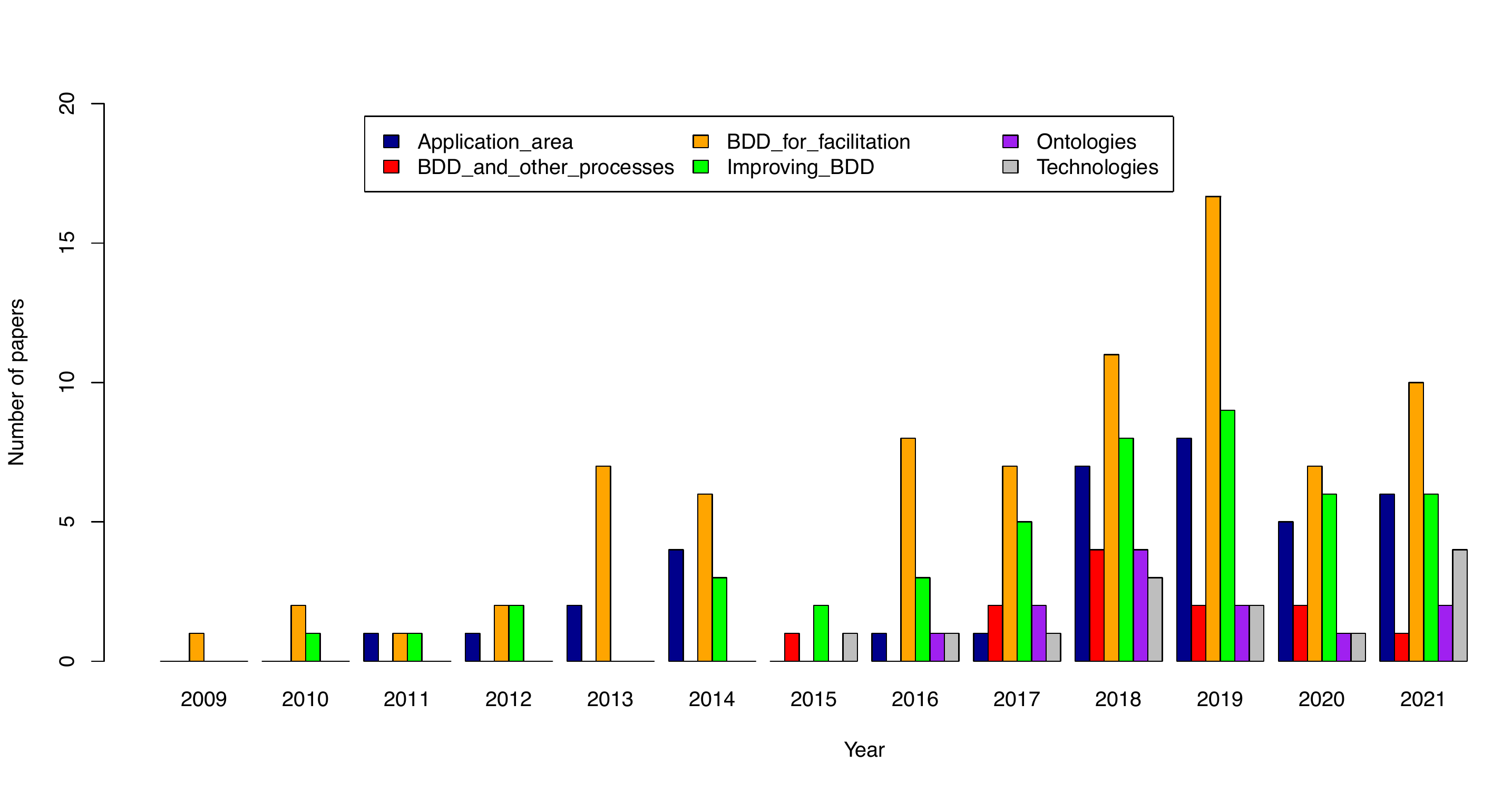}
   \vspace*{-3mm}
 \caption{Trend of themes over the years}
 \label{themes-over-years}
\end{figure}

Figure~\ref{themes-over-years} shows how the themes in BDD research (Table~\ref{existing-themes}) have changed over the years. In general, the number and diversity of studied topics in published BDD papers have been increasing over time. Similar to the number of papers, more diversity in the studied topics was mainly observed from 2016 to 2021, in which almost all themes were studied each year. More specifically, the results in Figure~\ref{themes-over-years} show that there has been a consistent and increasing interest in studying the use of BDD to facilitate various aspects of software engineering. This can be seen in all years but 2015. Generally, the papers on how BDD facilitates various aspects of software engineering have increased over time from 2009 to 2019, but seem to have reduced in 2020 and 2021. A similar trend is observed with studies on improving the actual BDD process. Studies on using BDD in different application areas have also increased over time and, on a deeper analysis, the diversity of domains has also increased. In 2015, we see new themes such as studying BDD alongside other software development processes and with different technologies. In 2016, the first paper on BDD and ontologies was published. However, the number of papers in these themes has not picked up over the years like papers in the other themes. 

\subsection{RQ8: Distribution of contributions w.r.t research types and topics}
\begin{figure}[!htb]
\centering
  \includegraphics[width=\textwidth]{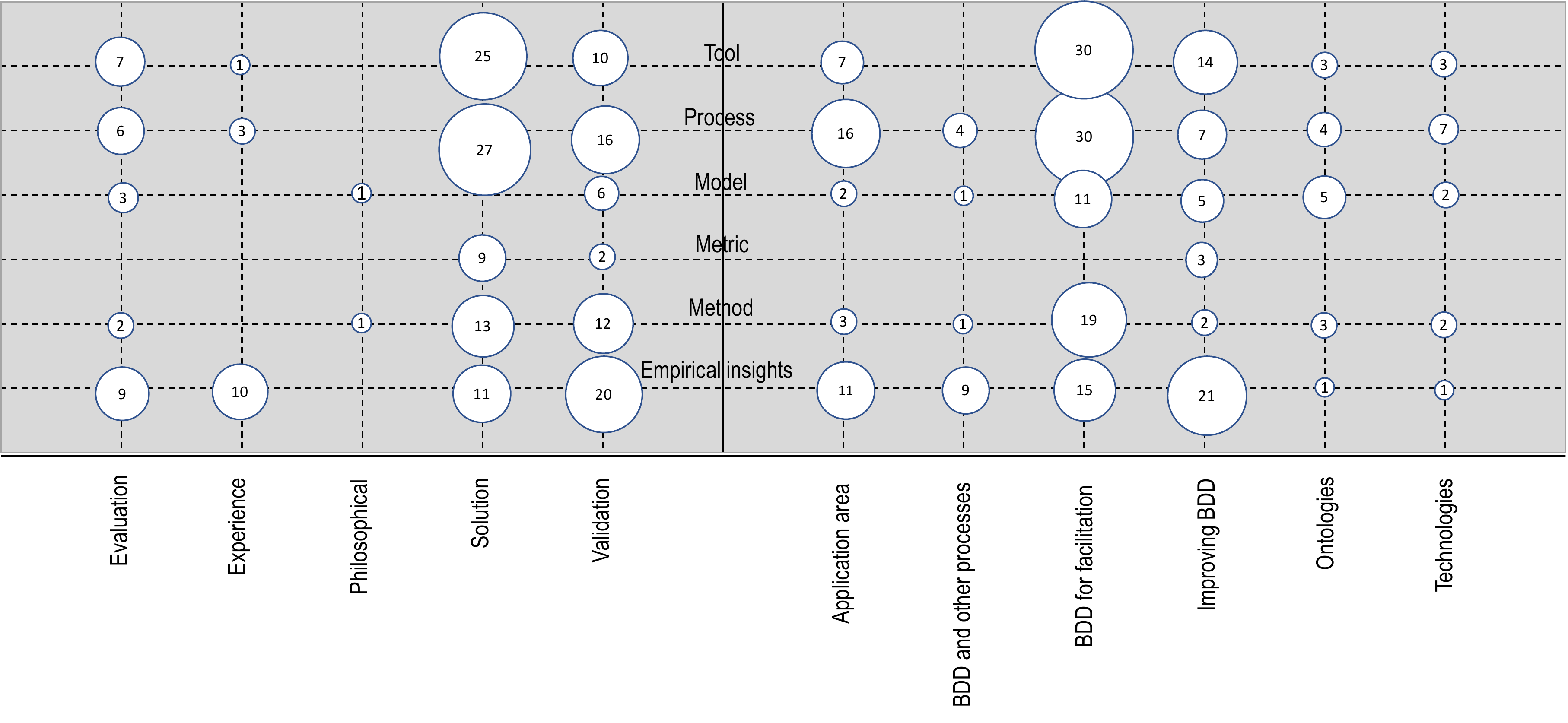}
   \vspace*{-3mm}
 \caption{Map of research types, themes, and contribution types}
 \label{systematic-map}
\end{figure}

Figure~\ref{systematic-map} shows the map that combines research types, contribution types, and themes in BDD research.

Generally, research on BDD has been dominated by studies in which the research type is solution proposal and the contribution type is process. Second in amount are solution proposals that have been accompanied by tools. Examples of these tools are tools for modelling scenarios~\cite{wiecher2021besos}, ensuring consistency between scenarios and GUI prototypes~\cite{silva2019extending} and code generation from scenarios~\cite{storer2019behave}. Apart from solution proposals, validation research is another research type that dominates existing research on BDD. Specifically, dominant contributions in the conducted validation research are of type process, empirical insights, method, and tool. Thus, there is a scarcity of models and metrics in the conducted validation research. Notably, there is a scarcity of BDD research that has been evaluated in industry settings, as well as philosophical and experience papers for all types of contributions.

Moreover, based on Figure~\ref{systematic-map}, the following can be noted regarding themes and contribution types. Most BDD studies have focused on producing tools, processes, methods, empirical insights, and models for facilitating software engineering processes. As regards improving the BDD process and the resulting artefacts, it can be seen that the contributions in most of the research on improving BDD are empirical insights and tools. Studies on the application of BDD in different settings have mostly contributed processes, empirical insights and tools, albeit to a relatively small extent. There is, however, a considerable lack of models, metrics and methods to inform the use of BDD in different settings. Also, most ontologies in existing BDD research have been accompanied by models, processes, methods, and tools. Nevertheless, the proposed ontologies have been accompanied by minimal empirical insights. This suggests the need for more BDD-related ontologies that are supported by strong empirical evidence. As well, when BDD was studied alongside other software engineering processes, the studies contributed more empirical insights and processes than models, methods and metrics. Finally, fewer processes, methods and tools were contributed when BDD was studied by integrating it with other technologies.


\section{Discussion}
\label{discussion-section}
\subsection{Publication venues for BDD papers} The publications about BDD are scattered in different journals, conferences and workshops. This suggests that BDD papers are not concentrated in specific journals, conferences or workshops, and implies that there are very few specific venues that publish BDD research.  The International Conference on Agile Software Development has published 7 papers over 10 years, making it one of the popular venues for BDD research. Moreover, our analysis shows that BDD research publication venues, even in the form of workshops, do not exist yet. 
Therefore, we call upon the software engineering research community to establish BDD-specific publication venues to consolidate the BDD research. Doing so will easily inform both the state-of-the-art and the state-of-the-practice.

Additionally, referring to Figure~\ref{distribution-of-publications-by-venue}, it can be seen that the majority of BDD research papers have been published in conference proceedings. Although there are very strong software engineering conferences such as the International Conference on Software Engineering (ICSE), generally, conference papers have space limitations, which restrict the amount of content in papers published in conference proceedings. The scarcity of journal papers means that most research on BDD has been published in brief as part of conference and workshop proceedings, which could lack substantial and elaborate insights of interest to both researchers and practitioners. As noted by Montesi and Lago \cite{montesi2008software}, journal publications allow extended discussion of research work with little worry about the limitations of space and time. We, therefore, encourage the software engineering research community to produce more journal publications with rich insights on BDD research to better inform research and practice.

\subsection{Researchers on BDD}
The results show that there are approximately 600 unique authors and co-authors working in the included set of papers. However, most of these researchers have only published a single paper or two in the area. The diverse nature of papers found by the present study can imply that researchers have other fields that they are working on and the BDD papers they have published have been opportunistic papers since BDD as a process is applicable in many domains. For instance, papers under the BDD application areas theme are published by researchers who are working on different domains and not necessarily software engineering. 

\subsection{Types of BDD research} The dominance of solution proposals and validation research over evaluation research (Figure~\ref{distribution-of-research-types} and Table~\ref{research-types-studies}) means that most of the BDD research either ended up proposing solutions for specific software engineering problems or went as far as a lab evaluation of the proposed solutions. 
The scarcity of studies that have been evaluated in the industry could affect the applicability of the proposed solutions because of the lack of evidence to demonstrate their effectiveness, when there is a general expectation that software engineering research should solve problems faced by practitioners \cite{pfleeger1994science}. 
Moreover, the fact that there are a few papers reporting the experience of using BDD in practice implies that more BDD experience reports are required, so that, the software engineering research community can understand BDD aspects that are important in practice, how the important BDD aspects are used, and the results they give \cite{klotins2019software}. This will, in turn, enable BDD researchers to prioritize aspects that matter to BDD practitioners. Thus, BDD practitioners need to report more on their experiences of working with BDD. In addition, more BDD philosophical papers are required to properly guide BDD research and practice. For example, papers on the theoretical foundations of BDD could enable the software engineering community to better understand how BDD is related to well-established software engineering theories. Finally, another lacking type of research is opinion papers. Since opinion papers present personal reflections on whether a certain method, technique or tool is good or bad, the lack of such studies means that the research community does not know how practitioners as well as academics feel about certain aspects of BDD. However, it is possible that such opinions are channelled through non-peer-reviewed venues such as blogs, because it is common for researchers and practitioners to air their opinions through blogs rather than scientific peer-reviewed venues.  Therefore, conducting multi-vocal literature reviews may be valuable to uncover personal opinions on the BDD topic.   

\subsection{Contribution types in BDD research} 
The results in Figure~\ref{distribution-of-contribution-types} and Table~\ref{contribution-types-studies} indicate that, while there is an arguably significant amount of other types of contributions, there is an acute shortage of studies on metrics for measuring various aspects of BDD specifications and the processes for producing BDD specifications. Out of the three studies that proposed metrics, two focus on measuring the quality of individual scenarios in a BDD specification \cite{oliveira2019evaluate, oliveira2017empirical}, and one focuses on measuring the overall quality of a BDD specification \cite{binamungu2020characterising}. However, there are no studies that have proposed metrics for measuring other aspects of the BDD technique and the resulting specifications--e.g., the quality of a BDD workflow used by a software team. The lack of studies on metrics on different aspects of BDD indicates that practitioners lack ways to measure the quality of the BDD process and its associated aspects, and ways to measure the impact and value that BDD brings. 

\subsection{Evaluation methods used in BDD research} Although some BDD researchers have employed rigorous evaluation methods such as case studies or experiments to study various aspects of BDD, most of the BDD studies have used example applications to demonstrate the applicability of solutions or insights generated in BDD research (Figure~\ref{research-methods} and Table~\ref{evaluation-methods-studies}). This implies that a substantial proportion of BDD research is immature, because, as observed by Chen and Babar \cite{chen2011systematic}, example applications, experience reports, and discussions are evaluation methods that cannot be considered to be scientifically rigorous. BDD researchers should, therefore, invest more efforts in rigorous empirical evaluation methods to increase the chances of transfer of research results into practice \cite{dyba2005evidence, chen2011systematic}. In addition, as mentioned in Section~\ref{conducted-research-section}, there is a paucity of studies reporting experiences of using BDD in industry, which limits the understanding of the software engineering research community on how BDD is used in practice in order to glean lessons that could inform future BDD research. Moreover, only three studies \cite{lucassen2017behavior, diepenbeck2018behaviour, diepenbeck2014behaviour} have performed rigorous analysis through formal methods. Specifically, formal methods in BDD have been used for hardware testing and verification \cite{diepenbeck2018behaviour, diepenbeck2014behaviour} and for tracing software requirements \cite{lucassen2017behavior}. More studies focusing on formal aspects of BDD are required to reap the benefits of formal methods in software engineering, which include producing properly abstracted and unambiguous software specifications \cite{plat1992application}.

\subsection{Evolution of BDD research over time} BDD research has evolved over time and space, producing more publications that cover a wide range of topics (Figure~\ref{frequency-of-publications-per-year} and Figure~\ref{themes-over-years}). This shows that various aspects of BDD have been attracting the attention of the software engineering research community. More specifically, initial research on BDD was about using BDD to facilitate software documentation \cite{brolund2009documentation}. However, over the years, apart from paying attention to how BDD facilitates other software development endeavours, BDD researchers have also focused on other topics (Figure~\ref{themes-over-years} and Table~\ref{existing-themes}). With respect to the application context, in 2011, BDD was first studied in the context of enterprise systems \cite{gohil2011towards}, but similar studies have been conducted to cover other specific contexts such as health \cite{hatko2014behaviour, purkayastha2020continuous, mello2018applicability, morrison2013proposing, morrison2013proposingnew, giorgi2019transition, bruschi2019behavior}, education \cite{Sarinho2019essemble, nascimento2020behavior, rocha2021enhancing, goulart2014using}, and others (Figure~\ref{themes-over-years} and Table~\ref{existing-themes}). This shows that BDD has a potential to be adopted in any domain where software is used. Moreover, studies on using BDD alongside other software development processes, techniques, and technologies started a bit later (2015 onwards), but the popularity of such studies is still relatively low compared to studies on other BDD topics (Figure~\ref{themes-over-years}). Among other things, this limits our ability to understand how the combination of BDD and other techniques and technologies such as scrum, microservices, machine learning, and blockchain can produce better solutions for different software engineering problems. Nevertheless, the noted evolution and diversity of the studied BDD topics over time suggest that more new and diverse BDD topics could be studied in the future, improving our understanding of the use of BDD in software development.

Importantly, apart from paying attention to other BDD topics discussed in the preceding paragraph, BDD researchers have had a sustained interest in improving the BDD process and the artefacts produced by the BDD process (Figure~\ref{themes-over-years}). This implies that the BDD process and the resulting artefacts could get better in the future, improving the way practitioners use BDD to develop and maintain software systems. However, as we indicate in Section~\ref{future-research-section}, further research on the studied topics is required to produce more mature insights to guide the work of BDD practitioners.

\subsection{Distribution of contributions w.r.t research types and topics}
The finding that BDD research has been dominated by studies in which the research type is solution proposal and the contribution type is process (Figure~\ref{systematic-map}) implies that most BDD studies have proposed processes as well as associated artefacts that act as solutions for software engineering problems. This is not a surprising finding due to the following explanation. As can be seen from many papers reported in the present study (see, for example, Figure~\ref{themes-distribution}), research on BDD has been focusing primarily on the artefacts (the BDD stories) resulting from the BDD process. One of the hypotheses could be that the BDD process is quite similar to the TDD process (which is already well known and studied), so the novelty brought by BDD lies more in the notation for specifying the scenarios (the artefacts), as it opens many opportunities for automation and support of other software development activities.

In addition to the paucity of specific types of contributions such as metrics, there is a notable scarcity of BDD research that has been evaluated in industry settings, as can be deduced from the small amount of evaluation research of all types of contributions. This leads to a lack of understanding of the applicability of the proposed solutions. Finally, the scarcity of philosophical and experience papers for all types of contributions could pose different challenges, including the following two. One, there is a possibility of having BDD research that is not properly guided (by appropriate theoretical foundations, which could be obtained from philosophical papers). Two, due to the scarcity of experience reports, in which lessons about the use of BDD in practice are shared, software teams could improperly practice BDD, failing to reap the envisaged benefits.

\section{Future Research}
\label{future-research-section}
In addition to the research opportunities that can be deduced from the discussion of results (Section~\ref{discussion-section}), we now discuss some key opportunities for BDD research, based on the results of the present study.

First, given a relatively small number of journal papers compared to conference papers (Figure~\ref{distribution-of-publications-by-venue}), there is a need for BDD researchers to focus more on producing mature and substantial studies, publishable by journals, to better inform the BDD state-of-art and state-of-practice. Although some software engineering conferences publish mature work, almost all of them have space limitations, hindering the ability of most authors to report in full and extensively. This might, in turn, limit the understanding and application of knowledge in conference and workshop papers. Although also limited, the space provided by software engineering journals is usually higher than the space provided by software engineering conferences. So, more journal publications about BDD would give a clear understanding of research achievements in the area.

Second, based on the scarcity of studies that have been evaluated in the industry (Figure~\ref{distribution-of-research-types} and Figure~\ref{systematic-map}), there is a need for more empirical evidence and insights which are informed by actual practices in the software industry. This will increase the chances of applicability of the proposed solutions because of reliable evidence to demonstrate their (solutions) effectiveness. Moreover, there is a need for more experience reports on BDD, to share best practices that can inform the user and research community about how BDD is actually used in the industry. In addition, future research should consider philosophical aspects of BDD, to better guide this study field. Among others things, philosophical studies could include theories, frameworks and taxonomies on various aspects of the BDD technique.

Third, based on the distribution of different contributions in existing BDD research (Figure~\ref{distribution-of-contribution-types} and Figure~\ref{systematic-map}), future studies should give due attention to metrics for measuring various aspects of BDD specifications and the processes for producing BDD specifications. Future research should also put more emphasis on developing models and methods to address different facets of BDD.

Fourth, the results in Figure~\ref{systematic-map} suggest the need for more processes, models, methods, and metrics that aim to improve the BDD process and associated artefacts. This is crucial for the development of the BDD research field since most studies have focused on the use of BDD in different contexts and to facilitate different activities, but few have focused on how core aspects of BDD such as scenario specifications can be improved. 
Fifth, the results in Figure~\ref{systematic-map} indicate that, apart from the need of developing more BDD-related ontologies, there is a need for BDD-related ontologies that are accompanied by empirical insights. This will influence their use in industry settings.

Sixth, there is a need for more research that studies BDD alongside other software processes and technologies (refer to Figure~\ref{systematic-map} and themes-over-years). Specifically, since BDD is an Agile technique, studies on how it can be used with specific Agile methods such as Scrum and Kanban will be interesting for practitioners who already use Agile. Additionally, studies on how BDD can be enhanced with emerging technologies such as machine learning are needed to generate lessons that could inform efficient software development.

\section{Threats to Validity}
\label{validity-threats-section}

\subsection{Identification of primary studies}
There could be biases in the identification of the primary studies, for two reasons: first, there is a chance that the search string could have excluded relevant studies; second, reviewers could be biased in excluding papers. To mitigate the effects of these threats, the search string (Section~\ref{search-section}) was formulated to be as generic as possible to avoid missing any BDD-related publications. Also, as described in Section~\ref{search-section}, snowballing was conducted to account for papers that might have been missed through a database search. However, even with snowballing, there is still a chance that some papers might have been missed in our search. This is tolerable because, as noted by Petersen \textit{et al.} \cite{petersen2015guidelines}, different from a systematic literature review, which aims to capture all appropriate evidence on a particular research question,  the aim of a systematic mapping study is to give a good general overview of a research area.

Moreover, since the majority of the papers were initially screened by individual researchers, there is a chance that relevant papers could have been excluded. To mitigate the effects of this threat, there was a pilot stage, in which 20 papers were reviewed by both researchers to agree on when to exclude or include papers. In addition, individual researchers could only mark a paper as excluded when they were completely sure that a paper meets a specific exclusion criterion; all papers with doubts were marked as ``maybe'' and were later discussed by both researchers in a workshop.  

\subsection{Data extraction and mapping}
To mitigate biases during data extraction and mapping, two strategies were used. First, to categorise a paper into a specific research type, contribution type, or research evaluation method, we used established classification schemes of Wieringa \textit{et al.}~\cite{wieringa2006requirements}, Mujtaba \textit{et al.}~\cite{mujtaba2008software}, and Chen and Babar ~\cite{chen2011systematic} respectively. Second, to categorise a paper to a specific theme, an inductive approach was used, where themes would emerge as reviewers proceeded with the categorisation process. With this approach, there is a chance that different reviewers would assign different themes to the same paper. To mitigate the effects of this threat, we conducted a pilot workshop, in which 50 papers were randomly selected and both reviewers did the categorisation together. Each emerging theme was thoroughly discussed for both reviewers to agree on papers that would fit that theme. The rest of the papers were categorised by individual reviewers; however, a workshop was held afterward to discuss papers on which reviewers had doubts regarding the themes to which they (reviewers) had categorised the papers, as well as to discuss and harmonise new themes that had emerged during individual work.

\section{Related work}
\label{related-work-section}
We found three secondary studies that are related to BDD. One of these studies has focused on BDD alongside TDD, while the other two studies have focused on BDD only. Abushama \textit{et al.} \cite{abushama2020effect} conducted a systematic literature review to understand the effect of TDD and BDD on project success by focusing on time, cost, and customer satisfaction. BDD was found to be more likely to achieve higher customer satisfaction than TDD, and TDD consumed more development time than BDD. However, both BDD and TDD were found to consume more cost and time than traditional approaches to software testing.

The first of the two secondary studies that focused on BDD only was reported in 2017 by Egbreghts \cite{egbreghts2017literature}, who conducted a systematic literature review in order to describe what BDD is and how it is used in practice. In 2020, Lillnor and He \cite{lillnor2020systematic} reported a systematic mapping study on BDD, which was conducted as part of a bachelor thesis. This is the second secondary study on BDD. However, both these studies were conducted as student projects and are only partially reported (only abstracts are accessible) in the grey literature; none of them is reported in peer-reviewed scientific literature. This poses concerns about the scientific rigour and coverage of these studies. We, thus, set out to conduct a systematic mapping study that can provide a reliable help to the scientific community in understanding the current state of research on BDD and the gaps in the existing literature, to inform future research on BDD. 

\section{Conclusion}
\label{conclusion-section}
BDD enables software teams to specify software requirements based on a structured natural language format, which facilitates understanding of software requirements among all software project stakeholders. The specifications in a natural language also act as test cases that can be executed to verify the behaviour of a software. Since the introduction of BDD about two decades ago, the software engineering research community has been investigating various aspects of BDD. However, there is a lack of secondary studies on BDD that have been published in peer-reviewed scientific outlets. This makes it difficult to understand the state of BDD research, as well as its strengths and gaps that could inform future BDD research.

This paper reports a Systematic Mapping Study in which 166 papers were analysed to understand the state of BDD research from 2006 to 2021. The study has revealed that most BDD research has focused on three things: using BDD to facilitate various aspects of software development; improving the BDD process and the resulting artefacts; and applying BDD in different application areas. Some of the notable gaps include scarcity of BDD research that is linked to practices in the software industry, scarcity of research in which BDD is studied alongside other software development processes and technologies, scarcity of philosophical papers on BDD, and acute shortage of metrics for measuring various aspects of BDD specifications and the processes for producing BDD specifications.

\bibliography{bdd}

\appendix \label{Appendix}
\section{Distribution of individual papers}
This section contains details on the classification of the different papers according to the research type, contribution type, evaluation method and theme.  

\begin{table*}[!htbp]
  \centering
  \caption{Studies for each research type}
\begin{tabular}{p{0.30\linewidth}  p{0.70\linewidth}}
\hline
{\bf Research Type} & {\bf Primary Studies} \\
\hline
Validation Research & \cite{silva2017behavior, esposito2018behavior, kannengiesser2020behaviour, mens2019method, xu2021requirement, guncan2021user, irshad2021adapting, oran2017analysing, dookhun2019assessing, santos2018automated, nguyen2020automated, malik2021automating, Sarinho2019essemble, carrera2014beast, storer2019behave, nascimento2020behavior, lucassen2017behavior, carter2016bhive, deng2021bmt, binamungu2020characterising, wang2018combiningicse, wang2018combining, mwakyanjala2020concurrent, purkayastha2020continuous, aghayi2021crowdsourced, zampetti2020demystifying, binamungu2018detecting, brolund2009documentation, silva2021empirical, rocha2021enhancing, rocha2019ensuring, rocha2020evaluating, silva2019extending, ma2019graph, santos2019improving, haser2016business, oliveira2017empirical, oliveira2018understanding, yang2019predicting, ma2018scenario, wiecher2020scenarios, lopes2021scrumontobdd, wang2018speed, almeida2016ss, pinto2019user, rocha2019using, elshandidy2021using, cisneros2018experimental, wanderley2015evaluation, pyshkin2012requirements, ali2019behavior, okolnychyi2016study, hoisl2014comparing, goulart2014using, wolde2021behavior}\\
\hline
Solution Proposal & \cite{kudo2019conceptual, silva2017formal, wanderley2012framework, dimanidis2018natural, rahman2015reusable, santos2015study, sivanandan2014agile, soeken2012assisted, schur2017augmented, contan2017automated, kamalakar2013automatically, zaeske2021behavior, diepenbeck2012behavior, nezhad2018behavior, shafiee2018behavior, diepenbeck2018behaviour, diepenbeck2014behaviour, lopez2014behaviour, hatko2014behaviour, lazuar2010behaviour, snook2018behaviour, elshandidy2019behaviour, wiecher2021besos, carter2016bhivenew, de2012business, manuaba2019combination, lai2014combining, sathawornwichit2012consistency, gupta2019creation, pandit2016distributed, Sarinho2019essemble, winkler2019efficient, schneider2018enabling, rocha2019ensuringnew, rocha2020ensuring, matula2018enterprise, baillon2010executable, fazzolino2019feature, fischer2019formal, gao2016generating, bonfanti2018generation, li2017gherkin, oliveira2019evaluate, carvalho2013implementing, schoeneman2013integrating, wiecher2021iterative, king2014legend, bagheri2013light, matula2018ontological, bussenot2018orchestration, silva2021parsing, morrison2013proposing, morrison2013proposingnew, matula2018reengineering, okubo2014security, orucc2016testing, silva2016testing, barus2019implementation, mahalakshmi2017theoretical, liao2017toward, o2018toward, hesenius2014towards, pang2021towards, diepenbeck2013towards, siqueira2017using, ali2019using, cauchi2016using, patkar2021interactive, lima2021acceptance, wolde2020behavior, raharjana2020tool, yen2021applying, tuglular2021behavior, wiecher2021integrated} \\
\hline
Evaluation Research & \cite{bunder2019model, bjarnason2016multi, lucena2017semantic, bjarnason2015industrial, alhaj2019approach, pereira2018behavior, nascimento2020behavior, schulz2019behavior, alferez2019bridging, souza2018improving, so2017intuitive, williams2020legislation, binamungu2018maintaining, nguyen2020automated, karagoz2017reproducing, li2016skyfire, zafar2021towards, bunder2019towards, alhaj2017using, bruschi2019behavior} \\
\hline
Experience Papers & \cite{fazzolino2018assessing, bezsmertnyi2020behavior, gil2016behavior, bahaweres2020behavior, scandaroli2019behavior, de2017combining, moult2020compliance, zaeske2020leveraging, lubke2016modeling, mello2018applicability, cavalcante2019behavior, giorgi2019transition, butler2019behaviour} \\
\hline
Philosophical Papers & \cite{solis2011study, gohil2011towards, lenka2018behavior} \\ \hline
Opinion papers & \cite{farago2020towards} \\
\hline
\end{tabular}
\label{research-types-studies}%
\end{table*}%

\begin{table*}[!htbp]
  \centering
  \caption{Studies for each contribution type}
\begin{tabular}{p{0.30\linewidth}  p{0.70\linewidth}}
\hline
{\bf Contribution Type} & {\bf Primary Studies} \\
\hline
     Model & \cite{silva2017behavior, kudo2019conceptual, silva2017formal, silva2016testing, lenka2018behavior, rahman2015reusable, lucena2017semantic, solis2011study, guncan2021user, shafiee2018behavior, elshandidy2019behaviour, carter2016bhive, lai2014combining, winkler2019efficient, matula2018enterprise, so2017intuitive, williams2020legislation, yang2019predicting, matula2018reengineering, wolde2021behavior} \\
\hline
   Process & \cite{esposito2018behavior, kannengiesser2020behaviour, bunder2019model, dimanidis2018natural, lucena2017semantic, santos2015study, guncan2021user, irshad2021adapting, alhaj2019approach, soeken2012assisted, diepenbeck2012behavior, diepenbeck2018behaviour, diepenbeck2014behaviour, lopez2014behaviour, hatko2014behaviour, snook2018behaviour, elshandidy2019behaviour, wiecher2021besos, carter2016bhivenew, de2012business, de2017combining, wang2018combiningicse, wang2018combining, sathawornwichit2012consistency, purkayastha2020continuous, aghayi2021crowdsourced, Sarinho2019essemble, schneider2018enabling, rocha2019ensuring, silva2019extending, fazzolino2019feature, carvalho2013implementing, wiecher2021iterative, zaeske2020leveraging, bagheri2013light, lubke2016modeling, bussenot2018orchestration, silva2021parsing, karagoz2017reproducing, wiecher2020scenarios, lopes2021scrumontobdd, almeida2016ss, silva2016testing, liao2017toward, o2018toward, zafar2021towards, pang2021towards, pinto2019user, alhaj2017using, elshandidy2021using, cauchi2016using, farago2020towards, ali2019behavior, lima2021acceptance, wolde2020behavior, bruschi2019behavior, wiecher2021integrated} \\
\hline
    Method & \cite{wanderley2012framework, mens2019method, xu2021requirement, schur2017augmented, malik2021automating, carrera2014beast, schulz2019behavior, lucassen2017behavior, carter2016bhive, deng2021bmt, alferez2019bridging, manuaba2019combination, mwakyanjala2020concurrent, gupta2019creation, binamungu2018detecting, pandit2016distributed, rocha2019ensuringnew, rocha2020ensuring, ma2019graph, matula2018ontological, ma2018scenario, okubo2014security, mahalakshmi2017theoretical, diepenbeck2013towards, siqueira2017using, gohil2011towards, tuglular2021behavior} \\
\hline
     Tool  & \cite{wanderley2012framework, bunder2019model, dimanidis2018natural, lucena2017semantic, soeken2012assisted, nguyen2020automated, contan2017automated, kamalakar2013automatically, Sarinho2019essemble, carrera2014beast, storer2019behave, nezhad2018behavior, lopez2014behaviour, lazuar2010behaviour, wiecher2021besos, carter2016bhive, moult2020compliance, sathawornwichit2012consistency, brolund2009documentation, Sarinho2019essemble, rocha2019ensuringnew, rocha2020ensuring, baillon2010executable, gao2016generating, bonfanti2018generation, li2017gherkin, santos2019improving, schoeneman2013integrating, wiecher2021iterative, king2014legend, nguyen2020automated, karagoz2017reproducing, li2016skyfire, wang2018speed, orucc2016testing, silva2016testing, liao2017toward, hesenius2014towards, zafar2021towards, bunder2019towards, pinto2019user, rocha2019using, patkar2021interactive, wanderley2015evaluation, raharjana2020tool} \\
\hline
    Metric & \cite{binamungu2020characterising, oliveira2019evaluate, oliveira2017empirical} \\
\hline
Empirical Insights & \cite{bjarnason2016multi, irshad2021adapting, sivanandan2014agile, bjarnason2015industrial, oran2017analysing, fazzolino2018assessing, dookhun2019assessing, santos2018automated, bezsmertnyi2020behavior, zaeske2021behavior, gil2016behavior, bahaweres2020behavior, scandaroli2019behavior, pereira2018behavior, nascimento2020behavior, lopez2014behaviour, binamungu2020characterising, manuaba2019combination, de2017combining, wang2018combining, zampetti2020demystifying, silva2021empirical, rocha2021enhancing, rocha2020evaluating, fischer2019formal, souza2018improving, so2017intuitive, haser2016business, williams2020legislation, binamungu2018maintaining, mello2018applicability, oliveira2018understanding, morrison2013proposing, morrison2013proposingnew, karagoz2017reproducing, wang2018speed, cavalcante2019behavior, barus2019implementation, diepenbeck2013towards, giorgi2019transition, elshandidy2021using, ali2019using, cisneros2018experimental, pyshkin2012requirements,okolnychyi2016study, hoisl2014comparing, goulart2014using, yen2021applying, butler2019behaviour} \\
\hline
\end{tabular} 
\label{contribution-types-studies}%
\end{table*}%

\begin{table*}[!htbp]
  \centering
  \caption{Studies that have employed specific evaluation methods}
\begin{tabular}{p{0.30\linewidth}  p{0.70\linewidth}}
\hline
{\bf Evaluation Method} & {\bf Primary Studies} \\
\hline
   Example & \cite{kannengiesser2020behaviour, kudo2019conceptual, silva2017formal, wanderley2012framework, dimanidis2018natural, rahman2015reusable, sivanandan2014agile, soeken2012assisted, schur2017augmented, contan2017automated, zaeske2021behavior, diepenbeck2012behavior, nezhad2018behavior, diepenbeck2018behaviour, diepenbeck2014behaviour, hatko2014behaviour, lazuar2010behaviour, snook2018behaviour, wiecher2021besos, carter2016bhivenew, de2012business, wang2018combiningicse, sathawornwichit2012consistency, gupta2019creation, pandit2016distributed, brolund2009documentation, Sarinho2019essemble, schneider2018enabling, morrison2013proposing, morrison2013proposingnew, matula2018enterprise, baillon2010executable, fischer2019formal, bonfanti2018generation, carvalho2013implementing, schoeneman2013integrating, wiecher2021iterative, king2014legend, matula2018ontological, bussenot2018orchestration, silva2021parsing, matula2018reengineering, wiecher2020scenarios, almeida2016ss, orucc2016testing, mahalakshmi2017theoretical, o2018toward, hesenius2014towards, zafar2021towards, pang2021towards, siqueira2017using, cauchi2016using, patkar2021interactive, lima2021acceptance, wolde2020behavior, wolde2021behavior, raharjana2020tool, yen2021applying, tuglular2021behavior, wiecher2021integrated} \\
\hline
Case Study & \cite{silva2017behavior, bjarnason2016multi, lucena2017semantic, guncan2021user, irshad2021adapting, bjarnason2015industrial, alhaj2019approach, nguyen2020automated, malik2021automating, carrera2014beast, pereira2018behavior, nascimento2020behavior, schulz2019behavior, lopez2014behaviour, carter2016bhive, alferez2019bridging, binamungu2020characterising, manuaba2019combination, mwakyanjala2020concurrent, purkayastha2020continuous, aghayi2021crowdsourced, zampetti2020demystifying, winkler2019efficient, rocha2021enhancing, rocha2019ensuring, rocha2019ensuringnew, rocha2020ensuring, rocha2020evaluating, silva2019extending, fazzolino2019feature, oliveira2019evaluate, souza2018improving, so2017intuitive, williams2020legislation, binamungu2018maintaining, nguyen2020automated, oliveira2017empirical, oliveira2018understanding, karagoz2017reproducing, okubo2014security, li2016skyfire, liao2017toward, diepenbeck2013towards, alhaj2017using, elshandidy2021using, ali2019using, ali2019behavior, goulart2014using, bruschi2019behavior} \\
\hline
Experiment & \cite{esposito2018behavior, mens2019method, bunder2019model, xu2021requirement, oran2017analysing, dookhun2019assessing, santos2018automated, kamalakar2013automatically, storer2019behave, deng2021bmt, wang2018combining, zampetti2020demystifying, binamungu2018detecting, silva2021empirical, gao2016generating, li2017gherkin, ma2019graph, santos2019improving, haser2016business, yang2019predicting, ma2018scenario, lopes2021scrumontobdd, wang2018speed, barus2019implementation, bunder2019towards, pinto2019user, rocha2019using, hoisl2014comparing} \\
\hline
Experience Report & \cite{fazzolino2018assessing, bezsmertnyi2020behavior, gil2016behavior, bahaweres2020behavior, scandaroli2019behavior, de2017combining, moult2020compliance, zaeske2020leveraging, lubke2016modeling, mello2018applicability, cavalcante2019behavior, gohil2011towards, giorgi2019transition, butler2019behaviour} \\
\hline
Discussion & \cite{santos2015study, solis2011study, contan2017automated, shafiee2018behavior, elshandidy2019behaviour, lai2014combining, bagheri2013light, silva2016testing, pyshkin2012requirements, farago2020towards, okolnychyi2016study} \\
\hline
Rigorous Analysis & \cite{lucassen2017behavior, diepenbeck2018behaviour, diepenbeck2014behaviour} \\
\hline
\end{tabular} 
\label{evaluation-methods-studies}%
\end{table*}%

\afterpage{
\begin{longtable}{p{0.02\linewidth}  p{0.2\linewidth}  p{0.30\linewidth}  p{0.33\linewidth}}
\caption{Themes in existing BDD research}
    \label{existing-themes}\\
\hline
 {\bf Sn} & {\bf Theme} & {\bf Sub-theme(s)} & {\bf Citation(s)} \\
\hline
\multicolumn{ 1}{r}{1} & \multicolumn{ 1}{l}{Facilitating software development} &     Safety & \cite{wang2018combiningicse, wang2018combining, baillon2010executable, wiecher2020scenarios, wang2018speed} \\
\multicolumn{ 1}{c}{} & \multicolumn{ 1}{c}{} & GUI testing & \cite{silva2017formal, bunder2019model, sivanandan2014agile, schur2017augmented, bahaweres2020behavior, rocha2019ensuring, rocha2019ensuringnew, bunder2019towards} \\
\multicolumn{ 1}{c}{} & \multicolumn{ 1}{c}{} & Artefact generation & \cite{wanderley2012framework, bunder2019model, dimanidis2018natural, soeken2012assisted, schur2017augmented, nguyen2020automated, kamalakar2013automatically, malik2021automating, carrera2014beast, storer2019behave, nezhad2018behavior, lazuar2010behaviour, snook2018behaviour, carter2016bhive, carter2016bhivenew, deng2021bmt, alferez2019bridging, gupta2019creation, pandit2016distributed, silva2021empirical, fazzolino2019feature, gao2016generating, bonfanti2018generation, schoeneman2013integrating, king2014legend, williams2020legislation, nguyen2020automated, li2016skyfire, mahalakshmi2017theoretical, o2018toward, diepenbeck2013towards, bunder2019towards, pinto2019user, siqueira2017using, raharjana2020tool, butler2019behaviour} \\
\multicolumn{ 1}{c}{} & \multicolumn{ 1}{c}{} & Product lines & \cite{elshandidy2019behaviour, bagheri2013light, elshandidy2021using, tuglular2021behavior} \\
\multicolumn{ 1}{c}{} & \multicolumn{ 1}{c}{} & Compliance testing & \cite{lopez2014behaviour, moult2020compliance, williams2020legislation, morrison2013proposing, morrison2013proposingnew, pang2021towards} \\
\multicolumn{ 1}{c}{} & \multicolumn{ 1}{c}{} & Traceability & \cite{lucassen2017behavior, silva2021parsing, yang2019predicting} \\
\multicolumn{ 1}{c}{} & \multicolumn{ 1}{c}{} & Model testing & \cite{mens2019method, snook2018behaviour, snook2021domain, silva2021empirical, rocha2019ensuring, silva2019extending, fischer2019formal} \\
\multicolumn{ 1}{c}{} & \multicolumn{ 1}{c}{} & Business process modelling & \cite{de2012business, carvalho2013implementing, lubke2016modeling, matula2018reengineering} \\
\multicolumn{ 1}{c}{} & \multicolumn{ 1}{c}{} & Security and privacy & \cite{lai2014combining, purkayastha2020continuous, okubo2014security} \\
\multicolumn{ 1}{c}{} & \multicolumn{ 1}{c}{} & Model-based testing & \cite{sivanandan2014agile, li2016skyfire} \\
\multicolumn{ 1}{c}{} & \multicolumn{ 1}{c}{} & Integration testing & \cite{purkayastha2020continuous, bussenot2018orchestration} \\
\multicolumn{ 1}{c}{} & \multicolumn{ 1}{c}{} & Load testing & \cite{schulz2019behavior} \\
\multicolumn{ 1}{c}{} & \multicolumn{ 1}{c}{} & Syntax highlighting & \cite{matula2018ontological} \\
\multicolumn{ 1}{c}{} & \multicolumn{ 1}{c}{} & Reproducing bugs & \cite{karagoz2017reproducing} \\
\multicolumn{ 1}{c}{} & \multicolumn{ 1}{c}{} & Spreadsheets testing & \cite{almeida2016ss} \\
\multicolumn{ 1}{c}{} & \multicolumn{ 1}{c}{} & Test selection  & \cite{xu2021requirement} \\
\multicolumn{ 1}{c}{} & \multicolumn{ 1}{c}{} & Usability testing  & \cite{guncan2021user} \\
\multicolumn{ 1}{c}{} & \multicolumn{ 1}{c}{} & Software documentation  & \cite{brolund2009documentation} \\
\hline
\multicolumn{ 1}{c}{2} & \multicolumn{ 1}{l}{Improving BDD} & Specification quality & \cite{binamungu2020characterising, oliveira2019evaluate, binamungu2018maintaining, oliveira2017empirical, oliveira2018understanding, binamungu2018detecting} \\
\multicolumn{ 1}{c}{} & \multicolumn{ 1}{c}{} & BDD adoption & \cite{irshad2021adapting, contan2017automated, bezsmertnyi2020behavior, gil2016behavior, scandaroli2019behavior, pereira2018behavior, shafiee2018behavior, nascimento2020behavior, zampetti2020demystifying, binamungu2018maintaining, mello2018applicability, cavalcante2019behavior, barus2019implementation, zampetti2020demystifying} \\
\multicolumn{ 1}{c}{} & \multicolumn{ 1}{c}{} & Co-evolution of artefacts & \cite{sathawornwichit2012consistency, zampetti2020demystifying, santos2019improving, yang2019predicting, rocha2019using} \\
\multicolumn{ 1}{c}{} & \multicolumn{ 1}{c}{} & Duplicate detection and refactoring & \cite{binamungu2018detecting} \\
\multicolumn{ 1}{c}{} & \multicolumn{ 1}{c}{} & BDD characteristics & \cite{solis2011study, farago2020towards}\\
\multicolumn{ 1}{c}{} & \multicolumn{ 1}{c}{} & Specification language & \cite{kudo2019conceptual, lucena2017semantic, lazuar2010behaviour, deng2021bmt, snook2021domain, rocha2020evaluating, li2017gherkin, haser2016business, king2014legend, hesenius2014towards, pang2021towards, bunder2019towards, patkar2021interactive, wanderley2015evaluation, hoisl2014comparing, gutierrez2017modelling} \\
\multicolumn{ 1}{c}{} & \multicolumn{ 1}{c}{} & Test cases as requirements & \cite{bjarnason2016multi, bjarnason2015industrial, elshandidy2021using} \\
\multicolumn{ 1}{c}{} & \multicolumn{ 1}{c}{} & BDD tools & \cite{okolnychyi2016study, pyshkin2012requirements} \\
\hline
\multicolumn{ 1}{c}{3} & \multicolumn{ 1}{l}{BDD application areas} & Hardware context & \cite{kannengiesser2020behaviour, alhaj2019approach, zaeske2021behavior, diepenbeck2012behavior, nezhad2018behavior, diepenbeck2018behaviour, diepenbeck2014behaviour, deng2021bmt, mwakyanjala2020concurrent, winkler2019efficient, bussenot2018orchestration, zafar2021towards, alhaj2017using, cauchi2016using, yen2021applying, wiecher2021integrated} \\
\multicolumn{ 1}{c}{} & \multicolumn{ 1}{c}{} & Health context & \cite{hatko2014behaviour, purkayastha2020continuous, mello2018applicability, morrison2013proposing, morrison2013proposingnew, giorgi2019transition, bruschi2019behavior} \\
\multicolumn{ 1}{c}{} & \multicolumn{ 1}{c}{} & Education context & \cite{Sarinho2019essemble, nascimento2020behavior, rocha2021enhancing, goulart2014using} \\
\multicolumn{ 1}{c}{} & \multicolumn{ 1}{c}{} & Mobile apps & \cite{nguyen2020automated, ali2019using, ali2019behavior} \\
\multicolumn{ 1}{c}{} & \multicolumn{ 1}{c}{} & Enterprise systems & \cite{fazzolino2018assessing, gohil2011towards} \\
\multicolumn{ 1}{c}{} & \multicolumn{ 1}{c}{} & BDD and computer networking & \cite{esposito2018behavior} \\
\multicolumn{ 1}{c}{} & \multicolumn{ 1}{c}{} & Product configuration systems & \cite{shafiee2018behavior} \\
\multicolumn{ 1}{c}{} & \multicolumn{ 1}{c}{} & Multi-agent systems  & \cite{carrera2014beast} \\
\hline
         4 & Ontologies & Ontologies &  \cite{silva2017behavior, rocha2019ensuring, rocha2019ensuringnew, rocha2020ensuring, matula2018enterprise, souza2018improving, matula2018ontological, silva2021parsing, matula2018reengineering, lopes2021scrumontobdd, silva2016testing} \\
\hline
\multicolumn{ 1}{c}{5} & \multicolumn{ 1}{l}{BDD and other processes} & Comparison with other techniques & \cite{santos2015study, oran2017analysing, dookhun2019assessing, santos2018automated, bezsmertnyi2020behavior, manuaba2019combination, cisneros2018experimental} \\
\multicolumn{ 1}{c}{} & \multicolumn{ 1}{c}{} & BDD and Scrum & \cite{de2017combining, souza2018improving, lopes2021scrumontobdd} \\
\multicolumn{ 1}{c}{} & \multicolumn{ 1}{c}{} & BDD and DevOps & \cite{zaeske2020leveraging, giorgi2019transition} \\
\hline
\multicolumn{ 1}{c}{6} & \multicolumn{ 1}{l}{BDD and other technologies} & Machine Learning & \cite{deng2021bmt, ma2018scenario, pang2021towards} \\
\multicolumn{ 1}{c}{} & \multicolumn{ 1}{c}{} & Blockchain &  \cite{liao2017toward} \\
\multicolumn{ 1}{c}{} & \multicolumn{ 1}{c}{} & Microservices & \cite{rahman2015reusable, aghayi2021crowdsourced, ma2019graph, ma2018scenario, lima2021acceptance} \\
           & \multicolumn{ 1}{c}{} & Web services & \cite{dimanidis2018natural, orucc2016testing, wolde2020behavior, wolde2021behavior} \\
\hline
\end{longtable}}

\end{document}